\documentclass[12pt,twoside,draftclsnofoot,onecolumn]{IEEEtran}

\ifCLASSINFOpdf
\else
\fi

\usepackage{graphicx,subfigure,array,amssymb,latexsym,amssymb,lipsum,algpseudocode,epsfig,float,epsf,amsmath,epstopdf,cite,tikz, float,algpascal,amsthm, bm, stackengine,multicol,microtype}
\usepackage{dblfloatfix}
\usepackage[linesnumbered,ruled,vlined]{algorithm2e}
\usepackage{cuted}
\DeclareGraphicsExtensions{.pdf,.png,.jpg,.eps}

\newcommand\myeq{\stackrel{\mathclap{\normalfont\mbox{(a)}}}{=}}

\newcommand\beq{\stackrel{\mathclap{\normalfont\mbox{(b)}}}{=}}

\newcommand\ceq{\stackrel{\mathclap{\normalfont\mbox{(c)}}}{=}}

\newcommand\deq{\stackrel{\mathclap{\normalfont\mbox{(d)}}}{=}}

\newcommand\eeq{\stackrel{\mathclap{\normalfont\mbox{(e)}}}{=}}

\newcommand\feq{\stackrel{\mathclap{\normalfont\mbox{(f)}}}{=}}

\makeatletter
\newcommand{\removelatexerror}{\let\@latex@error\@gobble}
\makeatother

\newtheorem{rem}{Remark}
\newtheorem{theo}{Theorem}

\newtheorem{lemma}{Lemma}

\SetKwProg{Fn}{Function}{}{}
\SetKwRepeat{Do}{do}{while}
\usepackage{booktabs}
\usepackage{multirow}
\usepackage{mathtools}
\DeclarePairedDelimiterX{\norm}[1]{\lVert}{\rVert}{#1}

\graphicspath {{/Simulation/Figure/}}

\begin{document}
%





\title{Joint AP Association and PCS Threshold Selection in Dense Full-duplex Wireless Networks}



%


\author{\IEEEauthorblockN{Phillip B. Oni and Steven D. Blostein}\\
\IEEEauthorblockA{Dept. of Electrical and Comp. Eng., 
Queen's University, Kingston, ON, Canada.\\
Email: \{phillip.oni, steven.blostein\}@queensu.ca}}


%



\maketitle

\begin{abstract}

\noindent Joint access point (AP) association and physical carrier sensing (PCS) threshold selection has the potential to improve the performance in high density wireless LANs (WLANs) under high contention, interference and self-interference (SI) limited transmissions. Using tools from stochastic geometry, user and AP locations are independent realizations of spatial point processes. Considering the inherent effects of the channel access protocol, the spatial density of throughput (SDT), which depends on channel access probability and coverage rate, is derived as the performance objective. Leveraging spatial statistics of the network, a throughput-utility maximization problem is formulated to seek AP association and PCS threshold selection policies that jointly maximize SDT. The AP association and the PCS threshold selection policies are derived analytically while an algorithm is proposed for numerical solution. Under simulated scenarios involving full-duplex (FD) nodes, optimizing AP association yields performance gains for low to high node density in large-scale wireless networks. Considering PCS threshold selection optimization jointly with AP association is shown to improve performance by effectively separating concurrent transmissions in space. It is shown that AP association in FD WLANs groups users into minimal contention domains and PCS threshold optimization reduces the interference domain of user groups for additional performance gains.

\end{abstract}

\makeatletter{\renewcommand*\@makefnmark{}
\makeatother}

\section{Introduction}

Dense deployment of wireless local area network (WLAN) access points (APs) to serve densely distributed users or stations (STAs) is expected to continue beyond fifth generation (B5G) of 3GPP, to interface with cellular 5G/B5G systems for cellular-WiFi data offloading \cite{offloading} supporting different use cases and requirements. More precisely, spatial densification (or small cells) in unlicensed spectrum could provide additional capacity for delivering best-effort and Internet of Things (IoT) traffic \cite{5garch}. However, high interference and contention from large numbers of concurrent spectrum-sharing nodes \cite{sandraL}, \cite{yingzhel} contribute to bottlenecks in scaling WLANs. Increasing AP and STA density increases the interference and contention domain of each node, thereby limiting throughput gain and spectral efficiency in high density WLANs. Although future enhancements to the physical layer (PHY) such as full-duplex (FD) transmission could potentially double capacity, high interference and contention may instead reduce the performance of FD communication.

Interference and contention in high density WLANs are tightly dependent on the density of concurrent spectrum sharing nodes. While interference results from large numbers of simultaneously active transmitters \cite{yicheng}, contention among nodes depends on the carrier sense multiple access collision avoidance (CSMA/CA) protocol \cite{sandraL} that governs access to the shared spectrum. These problems become aggravated in high density WLANs and thus necessitate mitigation techniques. Interference footprint and contention domains in the network depend on how users or STAs are distributed among the APs in terms of AP association, while the effectiveness of the CSMA/CA protocol to manage contentions from densely distributed users depends on the physical carrier sensing (PCS) threshold that spatially separates multiple concurrent transmissions. Current strongest-signal-first (SSF) association \cite{weili}, where users associate with the closest AP, performance does not explicitly take interference among the users into account. Similarly, the existing globally fixed PCS threshold may not guarantee successful transmission under the CSMA/CA protocol. 

Inevitably, dense WLAN (DWLAN) deployments will have multiple overlapped basic service sets (OBSSs), and thereby increase the interference domain of each AP \cite{shin}. Under the CSMA/CA protocol, OBSS could potentially reduce the number of concurrent pairs of FD transmissions in the network due to severe channel access contentions among nodes. To address these phenomena inherent in DWLAN, this paper investigates the joint user-AP association and PCS threshold selection problem in high density WLANs where FD transmissions are susceptible to interference, self-interference (SI) and CSMA/CA protocol effects (contention). Our main objective is to maximize the average spatial performance by defining the spatial density of throughput (SDT), which takes spatial statistics of the network into account, including node density, spatial topology, channel access probability of a typical node and its coverage likelihood, expressed in terms of the successful transmission probability (STP). We seek a solution that jointly associates users with the APs and spatially separates multiple concurrent transmissions via PCS threshold selection to improve spatial average throughput.

\subsection{Related Work}


Conventional user-AP association schemes, and the SSF scheme in particular, tend to ignore the interference and the contention level at the APs, and is based on AP association decisions mainly on the strongest received signal strength (RSS). Existing approaches to the user-AP association problem include load balancing techniques \cite{yzhangd, gAthan}, \cite{qyerong} that associate users to AP based on load metric, AP association maximizing proportional fairness\cite{weili}, and the decoupled user association (DUA) approach \cite{ahmed} that allows users to be served by different APs for uplink (UL) and downlink (DL) transmissions in FD WLANs. While optimizing AP association could improve performance in wireless networks, other lingering problems such as channel access protocol effects, spectrum allocation, power allocation, scheduling and fairness often cause performance degradation. To that effect, different approaches are proposed in the literature for joint AP association and spectrum allocation \cite{yinjunL, xiaofeng}, joint user-AP association and power allocation\cite{liyan, trinhvan, haijun}, user-AP association and user scheduling \cite{mengjie2, mengjie}, AP association and load balancing \cite{zhayang, leiyou, made, qyerong}, and joint AP association and proportional fairness \cite{gAthan, weili}.







The issues of user association and spectrum allocation in heterogeneous networks are coupled in \cite{yinjunL} where the joint problem is solved to maximize sum rate.  Posing the user-AP association problem as a classical assignment problem, a distributed auction algorithm is used in \cite{yuzheXu} to jointly optimize AP selection and relaying for optimal client-relay-AP association. In \cite{mengjie}, to maximize user utility for UL MU-MIMO WLANs an auction-based AP selection and STA scheduling framework is proposed. The auction-based AP association control algorithms \cite{mengjie2}, \cite{mengjie}, \cite{yuzheXu} could potentially suffer from lack of fairness. Formulating the user-AP association as one of utility maximization is able to account for fairness \cite{weili}, through joint user-AP association and power allocation \cite{liyan}, as well as adopting a game theory viewpoint \cite{bethanab}. The major challenge in dense wireless networks is not necessarily related to load balancing because there is usually a large number of APs to ensure coverage to all users. Using AP utilization as the AP association metric could lead to other deleterious effects such as high interference and high contention. 








Since interference and contention levels in WLANs depend on the density of concurrent transmitters, which are permitted by the underlying CSMA/CA protocol, AP association schemes that account for protocol effects are desirable. To improve the efficiency of the CSMA/CA protocol, existing schemes surveyed in \cite{cthorpe} advocate tuning of the PCS threshold to optimally determine the number of simultaneous transmitters per time slot (spatial reuse) \cite{lester}. For next-generation WLAN systems, static PCS threshold would be inefficient for \textit{densification} \cite{adra}, \cite{valk}. Prior to this work, an optimal PCS threshold selection scheme is proposed in \cite{pbopc} and \cite{psdb} for dense MIMO and SISO WLAN, respectively. Investigations therein show the achievable gain via optimizing PCS threshold. In \cite{valk}, transmit power and channel access rules in WLANs are dynamically adjusted using Basic Service Set (BSS) coloring information. Assuming channel knowledge, the PCS threshold could be jointly optimized with transmission rate \cite{yozhang} or with transmit power \cite{roslan} for performance gains in high density WLANs. By examining network information (such as RSS and perceived interference) contained in the PHY header, the PCS threshold is adjusted for spatial reuse in \cite{yoo}, \cite{yoo2}. Similarly, in \cite{chau}, PCS threshold can be defined using feedback from nearby transmissions. 

Under dynamic sensitivity control (DSC) in the IEEE 802.11ax standard, PCS threshold rules are defined to account randomness of node location and the channel access behavior for stochastic WLANs \cite{aijaz}. In \cite{megumi}, the performance of FD WLANs are analyzed assuming imperfect collision detection (CD). A more related approach is found in \cite{yenakim}, where the AP association problem is jointly considered with PCS threshold adjustment. While the schemes in \cite{valk}, \cite{yoo}, \cite{yoo2} require perfect interference estimation, those proposed in \cite{yozhang}, \cite{roslan}, \cite{yenakim} require frequent channel sounding to tune the PCS threshold. These solutions require real-time frequent channel sounding to obtain channel information, resulting in costly overhead in high density networks. The approaches in \cite{atzeni1}, \cite{atzeni} establish the performance gains of FD MIMO in terms of successful transmission probability and spectral efficiency under interference cancellation in FD small-cells wireless networks.

Most of the previous works described above either focus on user-AP association optimization or PCS threshold optimization. The approach taken in this paper is to jointly optimize both AP association and PCS threshold selection. This is motivated by the fact that optimal PCS threshold value depends on the distribution of users among the APs. In contrast to previous approaches, our objective is to establish a framework that improves AP association and enhances spatial reuse jointly without requiring detailed prior network information or channel sounding. Assuming prior knowledge of node density and distribution, the spatial average throughput of FD transmissions in MIMO WLAN is optimized by jointly optimizing AP association and PCS threshold selection. In contrast to the \textit{cut-through} FD mode \cite{sWang}, \cite{shuwang} for relay transmissions assumed in \cite{atzeni1}, \cite{atzeni}, our framework is based on the \textit{bidirectional} FD mode \cite{sWang}, \cite{shuwang} where APs and STAs transmit concurrently in both directions, with multiple antennas at both transmitter and receiver. Herein, the performance gains are obtained by optimizing user-AP association jointly with parameter tuning for spatial reuse.

\subsection{Contributions and Organization}

Herein, the primary objective is to efficiently perform AP association jointly with PCS threshold selection in high density MIMO full-duplex (FD) WLANs based on spatial statistics rather than on deterministic user-AP channels that require constant updates or \textit{a-priori} channel information. First, AP association is performed such that users (or STAs) are grouped in contention domains. Then the PCS threshold selection problem is solved for spatial reuse. To the best of our knowledge, despite large literature on AP association and PCS threshold selection problems, the AP association and PCS threshold selection problems are jointly considered here for the first time, and certainly for the case of MIMO FD WLANs with self-interference (SI). Our main contribution is summarized as follows.


\begin{itemize}
	\item Using tools from stochastic geometry we derive the mean rate utility termed \textit{spatial density of throughput} (SDT), which depends on the successful transmission probability (STP) and channel access probability in FD WLANs with SI. By maximizing the SDT, the optimal AP association policy is obtained along with the PCS threshold for optimal spatial reuse. Performance is evaluated in a MIMO FD WLAN. In particular, the throughput gain per unit area can be optimized by AP association, and by jointly considering PCS threshold, an additional throughput gain is obtained. 
	
	
	
\end{itemize}

The rest of this paper is organized as follows. The physical layer and network model assumptions are presented in Section~\ref{sysmodeljapo}. In Section~\ref{interferencemodel2}, interference and contention under CSMA/CA protocol is modeled and analyzed for both half-duplex (HD) and FD CSMA/CA networks, which determine a typical node's channel access probability, its successful transmission probability and the PCS threshold constraint. The performance metric, SDT is derived in Section~\ref{performmetric} while the proposed joint mean rate utility maximization framework is discussed in Section~\ref{framework}. Section~\ref{chjperformanceI} presents the performance evaluation and Section~\ref{conclusion} concludes this paper.

\section{System and Network Model}{\label{sysmodeljapo}}
\subsection{Network Model}

\begin{figure}[!h]
	\centering
	\includegraphics[width=4.5in, height=2.5in]{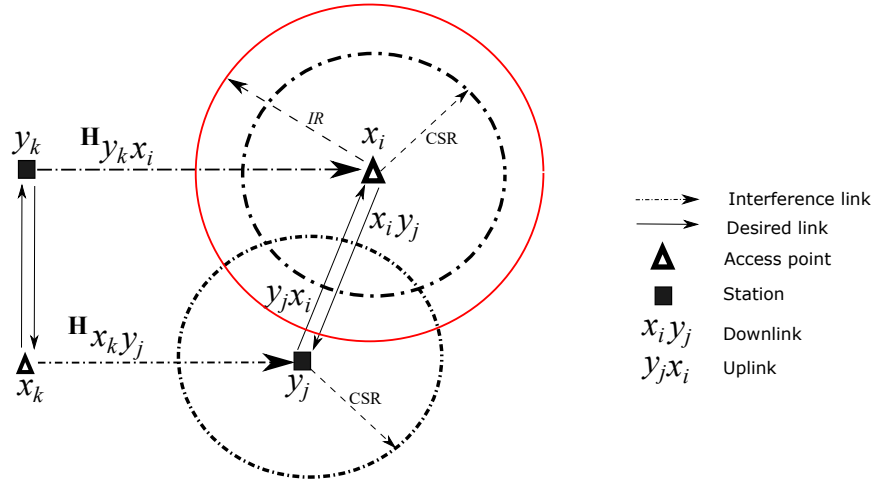}
	\caption{Bi-directional FD mode.}
	\label{fig12}
\end{figure}

We assume a multi-cell WLAN where the AP (or BSS) in each cell serves its associated STAs, one at a time in the presence of self-interference and out-of-cell interference. As shown in Figure~\ref{fig12}, assume that each AP initiates FD communication to serve the STA. We assume an unplanned AP deployment such as found in multi-tenant WLANs. We further assume that locations of APs and STAs follow independent realizations of Poisson point process (PPP). Let $\Phi_a = \{x_1, x_2, \cdots, x_{|\Phi_a|}\} \subset \mathbb{R}^2$ denote the distribution of APs with intensity $\lambda_a$. Similar to \cite{sWang}, \cite{shuwang}, where random locations of nodes are modeled using PPP, the locations of the STAs also follow a  homogeneous PPP $\Phi_s = \{y_1, y_2, \cdots, y_{|\Phi_s|}\} \subset \mathbb{R}^2$ with density $\lambda_s$. Assuming \textit{bi-directional} FD communication mode (discussed later in Section~\ref{fdmode2}) in a typical BSS, the received power for a pair of FD transmissions in each direction is:
\begin{equation}
\ell \left( x_i, y_j \right) = P^t\cdot\lVert  x_i - y_j \rVert^{-\alpha}, \qquad i = 1, \cdots, |\Phi_a|, j = 1, \cdots, |\Phi_s|
\label{pathloss}
\end{equation}

\noindent where $P^t$ is the fixed transmit power without power control, $\alpha$ is the path loss exponent, and $ \lVert  x_i - y_j \rVert $ denotes the \textit{Euclidean} distance between the \textit{primary transmitter} at point $x$ and the \textit{receiver} at point $y$. 

\subsection{Full-Duplex Communication Mode}{\label{fdmode2}}

In wireless LANs, full-duplex nodes operate in two modes, \textit{bidirectional transmission mode} and \textit{cut-through transmission mode} \cite{sWang}, \cite{shuwang}. When full-duplex nodes operate in \textit{bidirectional transmission mode}, an AP-STA pair is able to concurrently transmit to each other while the \textit{cut-through transmission mode} allows an access point (AP) to simultaneously transmit to two STAs; one uplink and one downlink transmission \cite{shuwang}. Herein, we will assume that the full-duplex wireless LAN allows only bidirectional transmission mode, as shown in Figure~\ref{fig12}, where the STA $y_j | j = 1, \cdots, |\Phi_s|$ and the AP $x_i | i = 1, \cdots, |\Phi_a|$ transmit concurrently to each other while AP $x_i$ and STA $y_j$ receivers experience interference from the STA $y_k | k = 1, \cdots, |\tilde{\Phi}_s|$ and AP $x_k | k = 1, \cdots, |\tilde{\Phi}_a|$, respectively, where $\tilde{\Phi}_s$ and $\tilde{\Phi}_a$ are the set of concurrently active STAs and APs, which are defined later in Section~\ref{interferencemodel2}. This interference model in \textit{bidirectional transmission mode} assumes the presence of SI and out-of-cell interference as discussed later in Section~\ref{interferencemodel2}.

\subsection{Physical Layer Model}

AP $x_i | i = 1, \cdots, |\Phi_a| $ and STA $y_j | j = 1, \cdots, |\Phi_s|$ are each equipped with transceivers having $M$ transmit antennas and $N$ receive antennas. From Fig.~\ref{fig12}, let $\mathbf{H}_{x_i y_j} \in \mathbb{C}^{N \times M}$ denote the channel from AP $x_i$ to STA $y_j$  and $\mathbf{H}_{y_j x_i} \in \mathbb{C}^{N \times M}$ denote the channel from STA $y_j$ to AP $x_i$. We assume that both $\mathbf{H}_{x_i y_j} \in \mathbb{C}^{N \times M}$ and $\mathbf{H}_{y_j x_i} \in \mathbb{C}^{N \times M}$ channels are Rayleigh fading with independent and identically distributed (i.i.d.) elements with zero mean and unit variance, i.e., $\mathcal{C}\mathcal{N}\left(0, 1\right)$. With imperfect self-interference (SI) cancellation in the system, the SI channel of STA $y_j$ and AP $x_i$ are denoted as $\mathbf{H}_{y_j y_j} \in \mathbb{C}^{N \times M}$ and $\mathbf{H}_{x_i x_i} \in \mathbb{C}^{N \times M}$, respectively. These SI channels are Rician fading with i.i.d. elements with mean $\mu$ and standard deviation $\psi^2$ \cite{mduarte}, that is, $\mathbf{H}_{y_j y_j} \in \mathbb{C}^{N \times M} \sim \mathcal{C}\mathcal{N}\left(\mu, \psi^2\right)$  and $\mathbf{H}_{x_i x_i} \in \mathbb{C}^{N \times M} \sim \mathcal{C}\mathcal{N}\left(\mu, \psi^2\right)$. The $K$-factor and the SI attenuation factor $\Omega$ \cite{ctepe}, are related to $\mu$ and $\psi^2$ of the SI channels via \cite{sgordon}:
\begin{equation}
\mu \triangleq \sqrt{\frac{K \Omega}{K + 1}} \mbox{ and } \psi^2 \triangleq \sqrt{\frac{\Omega}{K + 1}}.
\label{meanvarsi}
\end{equation}

Let $ \mathbf{V} \in \mathbb{C}^{M \times M} $ and $ \mathbf{W} \in \mathbb{C}^{N \times N} $ represent the transmit precoding and receive combining matrix at a typical transmitter and receiver, respectively. In the downlink (DL), the received signal before combining at the desired STA $y_j$ is given as:
\begin{equation}
\mathbf{y}_{y_j} =  \ell \left( x_i, y_j \right) \mathbf{H}_{x_i y_j}\mathbf{V}^{}_{x_i}\textbf{s}_{x_i} + \sum\limits_{x_k \in \tilde{\Phi}_a, i \neq k }^{  } \underbrace{\ell \left( x_k, y_j \right) \mathbf{H}_{x_k y_j} \mathbf{V}^{}_{x_k} \textbf{s}_{x_k} }_{I_o} + \underbrace{\ell\left( y_j, y_j \right) \mathbf{H}_{y_j y_j} \mathbf{V}^{}_{y_j} \textbf{s}_{y_j}}_{\mbox{SI}} + \mathbf{n}_{y_j},
\label{dlinkmodel}
\end{equation}
\noindent where $\ell \left( x_i, y_j \right)$ is the signal strength of the desired channel defined in Equation~(\ref{pathloss}), $\textbf{s}_{x_i}$ is the $M \times 1$ transmitted symbol vector from AP $x_i$ to STA $y_j$, $I_o$ is the out-of-cell interference from other transmitting APs $x_k$ in set $\tilde{\Phi}_a$ of concurrently active transmitters, $\ell\left( y_j, y_j \right)$ represents the signal strength of the SI channel and $\mathbf{n}_{y_j}$ is complex additive white Gaussian noise (AWGN) with zero mean and covariance, $\sigma^2 \mathbf{I}_{N}$. In the uplink (UL), the received signal at the AP $x_i$ is:
\begin{equation}
\mathbf{y}_{x_i} = \ell \left( y_j, x_i \right) \mathbf{H}_{y_j x_i}\mathbf{V}^{}_{y_j}\textbf{s}_{y_j} +  \sum\limits_{y_k \in \tilde{\Phi}_s, j \neq k }^{  } \underbrace{\ell \left( y_k, x_i \right) \mathbf{H}_{y_k x_i}  \mathbf{V}^{}_{y_k} \textbf{s}_{y_k} }_{I_o} + \underbrace{\ell\left( x_i, x_i \right)\mathbf{H}_{x_i x_i}  \mathbf{V}^{}_{x_i} \textbf{s}_{x_i}}_{\mbox{SI}} + \mathbf{n}_{x_i},
\label{uplinkmodel}
\end{equation}
\noindent where $\ell \left( y_j, x_i \right)$ is the signal strength from STA $y_j$ to AP $x_i$, $\textbf{s}_{y_j}$ is the uplink $M \times 1$ transmitted symbol vector from STA $y_j$, $I_o$ is the out-of-cell interference from other transmitting STAs $y_k$ in set $\tilde{\Phi}_s$ of concurrently active transmitters, $\mathbf{H}_{x_i x_i} \in \mathbb{C}^{N \times N}$ is the SI channel of AP $x_i$ and $\mathbf{n}_{x_i} \sim \mathcal{C}\mathcal{N}\left(0, \sigma^2 \mathbf{I}_{N}\right) $ is receiver noise at the AP. 

Considering the contention nature of the CSMA/CA protocol, the interfering node sets $\tilde{\Phi}_a$ and $\tilde{\Phi}_s$ are modeled later in Section~\ref{interferencemodel2}. Assuming channel side information (CSI) availability at both AP and STA, the received signals in Eqns.~(\ref{dlinkmodel}) and (\ref{uplinkmodel}) are processed via linear processing at the receivers considering interference from concurrent transmitters and SI. The DL signal-to-interference plus noise ratio (SINR) at STA $y_j$ after combining is
\begin{equation}
\mbox{SINR}_{y_j} = \frac{\xi_{x_i y_j}\ell \left( x_i, y_j \right) \lvert \mathbf{W}^{}_{y_j} \mathbf{H}_{x_i y_j}\mathbf{V}^{}_{x_i}\textbf{s}_{x_i} \rvert^2 }{ \norm{\mathbf{n}_{y_j}}^2 + \underbrace{\ell\left( y_j, y_j \right) \lvert \mathbf{W}^{}_{y_j} \mathbf{H}_{y_j y_j}\mathbf{V}^{}_{y_j}\textbf{s}_{y_j}\rvert^2}_{\mbox{SI}} + \underbrace{\sum\limits_{x_k \in \tilde{\Phi}_a, i \neq k }^{  } \ell \left( x_k, y_j \right) \lvert \mathbf{W}^{}_{y_j} \mathbf{H}_{x_k y_j} \mathbf{V}^{}_{x_k} \textbf{s}_{x_k}  \rvert^2}_{\mbox{out-of-cell interference}} },
\label{snrdl}
\end{equation}
\noindent where $\xi_{x_i y_j}$ is a binary variable, which indicates that STA $y_j$ is associated with AP $x_i$, i.e.,
\begin{equation}
\xi_{x_i y_j} = \left\{  \begin{array}{rcl}
1, & \mbox{if} \mbox{ STA } y_j \mbox{ associates with AP } x_i \\
0, & \mbox{otherwise.}  \qquad \qquad \qquad \qquad \qquad \qquad
\end{array}    \right.
\end{equation}

Similarly, from Eqn.~(\ref{uplinkmodel}), the uplink SINR at AP $x_i$ after combining via $\mathbf{W}^{}_{x_i}$ 

\begin{equation}
\mbox{SINR}_{x_i} = \frac{\xi_{x_i y_j}\ell \left( y_j, x_i \right) \lvert \mathbf{W}^{}_{x_i} \mathbf{H}_{y_j, x_i}\mathbf{V}^{}_{y_j}\textbf{s}_{y_j} \rvert^2 }{ \norm{\mathbf{n}_{x_i}}^2 + \underbrace{\ell\left( x_i, x_i \right) \lvert \mathbf{W}^{}_{x_i} \mathbf{H}_{x_i x_i}\mathbf{V}^{}_{x_i}\textbf{s}_{x_i} \rvert^2}_{\mbox{SI}} + \underbrace{\sum\limits_{y_k \in \tilde{\Phi}_s, j \neq k }^{  } \ell \left( y_k, x_i \right) \lvert \mathbf{W}^{}_{x_i} \mathbf{H}_{y_k x_i} \mathbf{V}^{}_{y_k} \textbf{s}_{y_k} \rvert^2}_{\mbox{out-of-cell interference}}  }.
\label{snrul}
\end{equation}

\noindent Equations (\ref{snrdl}) and (\ref{snrul}) are the SINRs of HD transmissions of downlink and uplink, respectively. The distributions of the desired signal powers $\lvert \mathbf{W}^{}_{y_j}\mathbf{H}_{x_i y_j}\mathbf{V}^{}_{x_i}\textbf{s}_{x_i} \rvert^2$ and $\lvert \mathbf{W}^{}_{x_i} \mathbf{H}_{y_j, x_i}\mathbf{V}^{}_{y_j}\textbf{s}_{y_j} \rvert^2$ in Eqs.~(\ref{snrdl}) and (\ref{snrul}), respectively, are Chi-square with $2M$ DoF \cite{goldsmith}. On the other hand, the SI powers $\lvert \mathbf{W}^{}_{y_j} \mathbf{H}_{y_j y_j}\mathbf{V}^{}_{y_j}\textbf{s}_{y_j}\rvert^2$ and $\lvert \mathbf{W}^{}_{x_i} \mathbf{H}_{x_i x_i}\mathbf{V}^{}_{x_i}\textbf{s}_{x_i} \rvert^2$ are characterized by a Gamma distribution with shape parameter $\kappa$ and scale parameter $\rho$ defined as \cite[Lemma 1]{atzeni}:
\begin{equation}
f\left(x\right) = \frac{x^{\kappa - 1} }{\Gamma\left(\kappa\right)\rho^{\kappa}} e^{-\frac{x}{\rho}}, \qquad \kappa \triangleq \frac{ \left( \mu^2 + \psi^2 \right)^2 }{ \Xi \mu^4 + 2 \mu^2 \psi^2 + \psi^4 }, \qquad \rho \triangleq \frac{ \Xi \mu^4 + 2\mu^2 \psi^2 + \psi^4 }{\mu^2 + \psi^2},
\label{siparams}
\end{equation}

\noindent where 
\begin{equation}
\Xi \triangleq \frac{4MN - \left(N + 1\right)\left(M + 1\right)}{\left(N + 1\right)\left(M + 1\right)}.
\end{equation}

\section{Interference and Contention Model under CSMA/CA Protocol}{\label{interferencemodel2}}

Under the carrier sense multiple access collision avoidance (CSMA/CA) protocol, each node competes for a transmission opportunity by sensing the channel for an active transmission. If no active transmission is detected during the physical carrier sensing (PCS) process, the node performing the PCS is allowed to transmit if the energy level detected on the channel is below a threshold $\Gamma$ known as the PCS threshold. The interference distribution on the network depends on the number of concurrent transmitters permitted by the CSMA/CA protocol following a contention period. In other words, a prospective transmitter senses the channel within its carrier sensing range (CSR) ${\mathcal{R}}$\footnote{CSR is the contention domain of each node.} and proceeds with its transmission if there is no other active transmitter within the CSR; two nodes will transmit concurrently if they are not within the same contention domain or CSR. To model this behavior of the CSMA/CA protocol to determine the set of concurrent transmitters\footnote{This set determines the amount of interference seen by a desired receiver.},  the Mat\`{e}rn hardcore point process (MHC PP)  (a.k.a Mat\`{e}rn Type II point process) \cite{sWang}, \cite{shuwang}, \cite{nguyen}, \cite{stoyan} is used. 

Considering the definition of MHC PP provided in \cite{nguyen}, \cite{stoyan}, a MHC PP of radius $\mbox{CSR}$ (\textit{see} Fig.\ref{fig12}) associated with homogeneous PPP $\Phi$ is obtained through a non-independent \textit{thinning} of $\Phi$. To perform the \textit{thinning} process, let us assign a uniformly distributed mark $m \sim U[0,1]$ to each point $x$ in $\Phi$. Then point $x$ will be selected if it has the lowest mark among other points within the radius, i.e., $m_{x} < m_{\bar{x}} \forall \bar{x}\in \Phi \setminus x$. The CSMA/CA permits node $x$ to transmit if the node has the minimum back-off time (or mark $m$) among other nodes in the contention region or domain $\mbox{CSR}$ and every other node within the contention domain will back-off while node $x$ completes its transmission. Although this model does not account for collisions, retransmissions and other delay effects of CSMA/CA protocol, it gives an accurate approximation of node transmission probability in dense WLAN scenarios \cite{nguyen}. In sections that follow, we describe this MHC PP model of the CSMA/CA protocol for HD and FD transmission modes.

\subsection{Half-Duplex (Uplink or Downlink) CSMA/CA}{\label{HDCSMA}}

Let $\tilde{\Phi}_{s}$ denotes the  Mat\`{e}rn thinning of the STAs PPP $\Phi_s$ and let $\tilde{\lambda}_s$ denote the intensity of $\tilde{\Phi}_{s}$. In set terms, $\tilde{\Phi}_{s} \subset \Phi$ is the set of concurrently transmitting STAs permitted by the CSMA/CA protocol. Using \textit{thinning process} discussed above to obtain $\tilde{\Phi}_{s}$ and $\tilde{\lambda}_s$, which defines the interference set in Eqn.~($\ref{uplinkmodel}$), we refer to Fig.~\ref{fig12}. A desired transmitter $y_j$ forms a circle $b\left(x_i, {\mathcal{R}} \right)$ with radius $\mathcal{R}$ as shown in Fig.~\ref{fig12} and contends with other transmitters within CSR ${\mathcal{R}}$. Let us associate a mark $m \sim U[0,1]$ to each contending STA within ${\mathcal{R}}$ centered at $y_j$. Transmitter $y_i$ will be retained in $\tilde{\Phi}_s$ if it has the lowest mark, i.e., $m_{y_j} < m_{\bar{y}_j} \forall \bar{y}_j \in \tilde{\Phi}^{y_j}_{\mathcal{R}} \setminus y_j$. In other words, this means that transmitter $y_j$ has the lowest CSMA backoff counter among all the transmitters within its CSR. $\tilde{\Phi}_s$ is the new PPP obtained from \textit{thinning} the parent PPP $\Phi_s$ and $\tilde{\Phi}^{y_j}_{\mathcal{R}} \subset \Phi_a$ is the set of transmitters that lie within the contention neighborhood of $y_j$. 

Supposed there are two transmitters $y_j$ and $\tilde{y}_j$ within ${\mathcal{R}}$, and the PCS threshold that determines the CSR ${\mathcal{R}}$ is given as $\Gamma$. Then $y_j$ and $\tilde{y}_j$ will transmit concurrently if $\ell \left( y_j, \tilde{y}_j \right)\norm{\mathbf{H}_{y_j \tilde{y}_j}}^2 < \Gamma$ and $\ell \left( \tilde{y}_j, y_j\right) \norm{\mathbf{H}_{\tilde{y}_j, y_j } }^2 < \Gamma$. This implies that the energy detected by each node during the carrier sensing is below threshold $\Gamma$; i.e., $y_j$ and $ \tilde{y}_j$ are each far enough apart to have a successful transmission with their respective receiver APs. Therefore, the \textit{Palm probability of retaining} $y_i$ in  $\tilde{\Phi}_s$ \cite{stoyan}, that is, the transmission probability that $y_j$ having the lowest backoff time is permitted by the CSMA/CA protocol and is given by
\begin{align}
p_{y_j} & = \int_{0}^{1} \mathbb{P}\bigg\{ y_j \in \tilde{\Phi}_s \bigg|  m_{y_j} = t \bigg\} \mbox{d}t = \frac{1 - \exp\left( - \lambda_s \Theta_s \right)}{ \lambda_s \Theta_s }
\label{probofy}
\end{align} 
\noindent where $\mathbb{P}\{ y_j \in \tilde{\Phi}_s |  m_{y_j} = t \}$ represents the probability of retaining a mark $t$ (or a backoff time value) as the lowest among other marks, which in terms of signal detection threshold during the PCS process, can be expressed as
\begin{align}
\mathbb{P}\bigg\{ y_j \in \tilde{\Phi}_s \bigg|  m_{y_j} = t \bigg\} & = \mathbb{P}\bigg\{ y_j \in \tilde{\Phi}_s \bigg|  m_{y_j} < m_{\bar{y}_j} \forall \bar{y}_j \in \tilde{\Phi}^{y_j}_{\mathcal{R}} \setminus y_j \bigg\} \\
& = \mathbb{P} \bigg\{ \ell \left( y_j, \tilde{y}_j \right)\cdot \norm{\mathbf{H}_{y_j \tilde{y}_j}}^2 < \Gamma \bigg\}
\end{align}
\begin{align}
& = 1 - \exp\left( - \frac{\Gamma}{ \ell \left( y_j, \tilde{y}_j \right) } \right) = 1 - \exp\left( - \Gamma \norm{ y_j - \tilde{y}_j}^\alpha   \right) ,
\label{probretain}
\end{align}
\noindent which follows from using the exponential property of the Chi-Square distribution with $2N$ degrees of freedom \cite{atzeni} of $\norm{\mathbf{H}_{y_j \tilde{y}_j}}^2 = \mbox{Tr}\left( \mathbf{H}_{y_j \tilde{y}_j} \mathbf{H}^H_{y_j \tilde{y}_j} \right)$, the channel power between $y_j$ and a contender $\tilde{y}_j$. Given volume $b_2$ of a unit ball in $\mathbb{R}^2$, $\Theta_s = b_2 \cdot \mathcal{R}^2$ is a volume integral over polar coordinates and can be written as
\begin{align}
\Theta_s & = 2\pi \int_{\mathbb{R}^+}^{} 1 - \mathbb{P}\bigg\{ y_j \in \tilde{\Phi}_s \bigg|  m_{y_j} = t \bigg\} \cdot \norm{ y_j - \tilde{y}_j} \mbox{ d} \norm{ y_j - \tilde{y}_j} \\
& = 2\pi \int_{\mathbb{R}^+}^{} \exp\left( - \Gamma \norm{ y_j - \tilde{y}_j}^\alpha   \right) \cdot \norm{ y_j - \tilde{y}_j} \mbox{ d} \norm{ y_j - \tilde{y}_j}.
\label{teta}
\end{align}

\noindent Without loss of generality, using \cite[Eqn~2.33.16]{tableInt}, Eqn.~(\ref{teta}) is simplified to
\begin{equation}
\Theta_s = \pi  \sqrt{\frac{\pi}{\Gamma}} \mbox{erf}\left(\sqrt{\Gamma}  \ell\left( y_j, \tilde{y}_j \right)\right).
\label{tetaval}
\end{equation}
\noindent where $\mbox{erf}\left( \cdot \right)$ is the error function and $\ell\left( y_j, \tilde{y}_j \right)$ is defined in Eqn.~(\ref{pathloss}) in terms of path loss exponent $\alpha$.

Hence, applying \cite[{Eqn. 5.55}]{stoyan}, the density of the resultant CSMA MHC PP $\tilde{\Phi}_s$ of the concurrently transmitting STAs permitted by the protocol becomes
\begin{equation}
\tilde{\lambda}_{s} = p_{y_j} \lambda_s =\frac{1 - \exp\left( - \lambda_s \Theta_s \right)}{  \Theta_s }.
\label{activesta}
\end{equation}

\noindent By substituting Eqn.~(\ref{tetaval}) into Eqn.~(\ref{activesta}), given a PCS threshold value $\Gamma$ and capturing the distance between a desired node $y_j$ and a potential interference source $\tilde{y}_j$, the density of active or concurrently transmitting STAs is expressed as
\begin{equation}
\tilde{\lambda}_{s}  =\frac{1 - \exp\left( - \lambda_s \pi \sqrt{\frac{\pi}{\Gamma}} \mbox{erf}\left(\sqrt{\Gamma}  \ell\left( y_j, \tilde{y}_j \right)\right) \right)}{  \pi \sqrt{\frac{\pi}{\Gamma}} \mbox{erf}\left(\sqrt{\Gamma}  \ell\left( y_j, \tilde{y}_j \right)\right) }.
\label{activesta2}
\end{equation}
\noindent By applying the same \textit{thinning} process, the PPP of the concurrent APs $\tilde{\Phi}_a$ similarly has the following density:
\begin{equation}
\tilde{\lambda}_{a} = p_{x_i} \lambda_a = \frac{1 - \exp\left( - \lambda_a \Theta_a \right)}{  \Theta_a } = \frac{1 - \exp\left( - \lambda_a \pi \sqrt{\frac{\pi}{\Gamma}} \mbox{erf}\left(\sqrt{\Gamma}  \ell\left( x_i, \tilde{x}_i \right)\right) \right)}{ \pi \sqrt{\frac{\pi}{\Gamma}} \mbox{erf}\left(\sqrt{\Gamma}  \ell\left( x_i, \tilde{x}_i \right)\right) }.
\label{activeap}
\end{equation}

Since the goal is to improve the spatial average of performance through user-AP association distribution, the average path loss $\ell\left( y_j, \tilde{y}_j \right)$ and $\ell\left( x_i, \tilde{x}_i \right)$ in Eqns.~(\ref{activesta2}) and (\ref{activeap}), respectively, need to be computed using a distance probability distribution. For STAs within the CSR $\mathcal{R}$, the distance $\norm{ y_j - \tilde{y}_j}$ between STA $y_j$ and its contending neighbor $\tilde{y}_j$ has a probability distribution characterized as

\begin{lemma}
	\label{ssflemma}
	The Euclidean distance $\norm{ y_j - \tilde{y}_j} \leq \mathcal{R}$ between STA $y_j$ and the nearest $n$th STA $\tilde{y}_j$ has a probability distribution given by:
	\begin{equation}
	f\left( \norm{ y_j - \tilde{y}_j} \right) = \frac{\left(2{\lambda}_{s}\pi\mathcal{R}^2\right)^n}{\mathcal{R}\beta(n)} e^{-{\lambda}_{s}\pi\mathcal{R}^2}.
	\label{ssfas}
	\end{equation}
\end{lemma}
\noindent \textit{Proof}. Since it is assumed that STA $y_j$ is at the origin of a 2-D ball with radius $\mathcal{R}$ to point $\tilde{y}_j$, the rest of the proof follows from \cite[Theorem 1]{mhag}.  \qed \\

\noindent If STA $y_j$ contends with only one STA $\tilde{y}_j$, the separation distance $\norm{ y_j - \tilde{y}_j}$ is Rayleigh distributed with expected value
\begin{equation}
\mathbb{E} \left[\norm{ y_j - \tilde{y}_j}\right] = \frac{1}{\lambda_{s} \pi},
\label{meandistanceSTA}
\end{equation}
\noindent and consequently,
\begin{equation}
\ell\left( y_j, \tilde{y}_j \right) = \left(\frac{1}{\lambda_{s} \pi}\right)^{-\alpha}.
\end{equation}
\noindent Similarly for the contending APs, the average path loss is
\begin{equation}
\ell\left( x_i, \tilde{x}_i \right) = \left(\frac{1}{\lambda_{a} \pi}\right)^{-\alpha}.
\end{equation}

\subsection{Full-Duplex (Bidirectional) CSMA/CA}{\label{FDCSMACA}}

In Section~\ref{HDCSMA}, the CSMA/CA model only considers winning contention for a single node among other nodes within the CSR. For a FD transmission to occur, the two nodes (in this case, one AP and one STA) must be permitted by the CSMA/CA protocol in the same time-slot. That is, a successful FD transmission depends on the probability of STA $y_j$ and AP $x_i$ being permitted concurrently by the CSMA/CA in the same time-slot. It is also assumed that nodes could switch between HD and FD transmissions depending on whether both nodes are granted transmission opportunity at the same time or not; that is, they transmit in FD mode if they both have access to the channel. Otherwise, either of the nodes operates in HD mode. 

By the \textit{superposition} principle \cite{stoyan}, given the two independent PPPs $\Phi_s$ and $\Phi_a$ of STAs and APs, respectively, for our FD network let $\Phi_{\mbox{FD}}$ denote the combined PPP of $\Phi_s$ and $\Phi_a$. The combined PPP has intensity
\begin{equation}
\lambda_{\mbox{FD}} = \lambda_s + \lambda_a.
\label{fddens}
\end{equation}
\noindent Therefore, using the same \textit{thinning} process in Section~\ref{HDCSMA}, the probability $p^{FD}_{x_i y_j}$ of retaining both nodes $y_{j}$ and $x_{i}$ in $\Phi_{\mbox{FD}}$ can be determined. In protocol terms, this probability $p^{FD}_{x_i y_j}$ represents the probability of FD transmission involving AP $x_i$ and STA $y_j$ being granted simultaneous access by the CSMA/CA protocol. Without loss of generality, considering the independence of $\Phi_s$ and $\Phi_a$, and their independent \textit{thinning} processes
\begin{eqnarray}
p^{FD}_{x_i y_j} =  \frac{1 - \exp\left( - \lambda_{\mbox{FD}} \pi \sqrt{\frac{\pi}{\Gamma}} \mbox{erf}\left(\sqrt{\Gamma}  \left( \lambda_{\mbox{FD}} \pi \right)^\alpha \right) \right)}{ \lambda_{\mbox{FD}} \pi \sqrt{\frac{\pi}{\Gamma}} \mbox{erf}\left(\sqrt{\Gamma}  \left( \lambda_{\mbox{FD}} \pi \right)^\alpha\right) }.
\label{fdprocap}
\end{eqnarray}

For the FD CSMA network, when a typical AP and its STA jointly win the contention period in the same time slot and activate the FD mode, the density of the FD transmissions in the network is given as
\begin{lemma}
	\label{denFD}
	The density of concurrent FD transmissions in the network is
	\begin{equation}
	\tilde{\lambda}_{\mbox{FD}} = p^{FD}_{x_i y_j} \cdot \lambda_{\mbox{FD}} = \frac{1 - \exp\left( - \lambda_{\mbox{FD}} \pi \sqrt{\frac{\pi}{\Gamma}} \mbox{erf}\left(\sqrt{\Gamma}  \left( \lambda_{\mbox{FD}} \pi \right)^\alpha \right) \right)}{ \pi \sqrt{\frac{\pi}{\Gamma}} \mbox{erf}\left(\sqrt{\Gamma}  \left( \lambda_{\mbox{FD}} \pi \right)^\alpha\right) }.
	\label{fddensity}
	\end{equation}
\end{lemma}
\noindent \textit{Proof}. The density of FD transmission pairs is the joint density of active APs with density $\tilde{\lambda}_{a}$ and the density of active STAs, $\tilde{\lambda}_{s}$, in Eqs.~(\ref{activeap}) and (\ref{activesta2}), respectively. Therefore, by multiplying the joint density $\lambda_{\mbox{FD}}$ of the PPPs $\Phi_s$ and $\Phi_a$ by the probability of FD transmissions in Eqn.~(\ref{fdprocap}), Eqn.~(\ref{fddensity}) is obtained. This yields the proportion $\tilde{\lambda}_{\mbox{FD}}$ of FD links permitted by CSMA/CA protocol for FD transmissions. \qed

%

\subsection{PCS Threshold Constraint}

Consider the system model in Figure~\ref{fig12}, the dotted circles represent the CSR, which is a guard zone where it is forbidden for two transmitters to transmit concurrently. The CSR determines spatial reuse and it is important to efficiently separate multiple concurrent transmissions in space. From Eqn.~(\ref{pathloss}), the CSR depends on the PCS threshold via path loss as
\begin{equation}
\Gamma = \mbox{CSR}^{-\alpha},
\label{pcsc1}
\end{equation}
\noindent and the PCS threshold should also be chosen to permit simultaneous spatially separated transmissions while keeping UL SINR above a threshold, $\gamma$. Assuming wireless links in Figure~\ref{fig12} are statistically \textit{homogeneous} in terms of interference and noise levels, The CSR can be enlarged to cover the entire potential interference range $IR$ (solid circle in Figure~\ref{fig12}) as:
\begin{align}
\mbox{CSR} & \geq \xi_{x_i y_j}\lVert  x_i - y_j \rVert + IR \\
& \geq \xi_{x_i y_j}\lVert  x_i - y_j \rVert \left(1 + P^t \gamma^{\frac{1}{\alpha}}\right) 
\label{csrr}
\end{align}
\noindent where $\mbox{IR}$ is defined as a function of the transmission range $\lVert  x_i - y_j \rVert$ and SINR threshold $\gamma$. Consequently, the constraint in Eqn.~(\ref{pcsc1}) can be written as
\begin{equation}
\Gamma \leq \xi_{x_i y_j} \lVert  x_i - y_j \rVert^{-\alpha} \frac{1}{ \left(1 + P^t \gamma^{\frac{1}{\alpha}}\right)^{\alpha} }.
\label{pcsc2}
\end{equation}

In each time slot, the above constraint on the PCS threshold determines the density of active nodes given in Equations~(\ref{activesta2}) and (\ref{activeap}), and consequently, the interference level in the network. In the next section, $\tilde{\lambda}_{s}$ and $\tilde{\lambda}_{a}$ derived in Eqns.~(\ref{activesta2}) and (\ref{activeap}) are applied to modeling concurrent transmitter sets $\tilde{\Phi}_s$ and $\tilde{\Phi}_a$, respectively, which are then used to compute successful transmission probabilities taking SINR into account.


\section{Performance in High Density WLANs}{\label{performmetric}}

A key performance metric in future high density WLAN is the density (or the number) of successful transmissions or spatial density of throughput (SDT). The throughput density is an indicator that represents the network performance per unit area, and is a function of the density of active nodes granted access by the CSMA/CA protocol (discussed in Section~\ref{interferencemodel2}) and the successful transmission probability (STP). STP is the probability that a node achieves a target SINR, which indicates successful signal reception at the receiver \cite{ykim}. Having derived the densities of active transmissions in both HD and FD cases in Section~\ref{interferencemodel2}, this section derives the STP in Section~\ref{transmprob} and formulates the SDT in Section~\ref{fdthruput}. The SDT is a suitable performance metric because it captures both the magnitude of interference from simultaneously active transmitters ($\tilde{\Phi}_s$ and $\tilde{\Phi}_a$) and the numbers of successful transmissions.

\subsection{Successful Transmission Probability (STP)}{\label{transmprob}}

 
A transmission is successful if the received SINR at the receiver is above a threshold $\gamma$. Under the bidirectional FD mode described in Section~\ref{FDCSMACA}, the AP receiver in the uplink (UL) and the STA receiver in the downlink (DL) experience different interference power levels. This is expected because the number of interferers in the UL is not usually the same as that in the DL, and \textit{channel reciprocity} does not hold. That is, the UL interference depends on set $\tilde{\Phi}_s$ while the DL interference depends on set $\tilde{\Phi}_a$ of concurrently transmitting APs. In fact, the probability of successful transmission of the DL is independent of the probability of successful transmission of the UL. For the UL signal received at the AP, we have  from (\ref{snrul})
\begin{align}
& \mathbb{P}\left(\mbox{SINR}_{{{x_i}}} \geq \gamma \right)  = \nonumber\\
&  \frac{\xi_{x_i y_j}\ell \left( y_j, x_i \right)\lvert \mathbf{W}^{}_{x_i} \mathbf{H}_{y_j, x_i}\mathbf{V}^{}_{y_j}\textbf{s}_{y_j} \rvert^2 }{ \norm{\mathbf{n}_{x_i}}^2 + {\ell\left( x_i, x_i \right) \lvert \mathbf{W}^{}_{x_i} \mathbf{H}_{x_i x_i}\mathbf{V}^{}_{x_i}\textbf{s}_{x_i} \rvert^2} + {\sum\limits_{y_k \in \tilde{\Phi}_s, j \neq k }^{  } \ell \left( y_k, x_i \right)  \lvert \mathbf{W}^{}_{x_i} \mathbf{H}_{y_k x_i} \mathbf{V}^{}_{y_k} \textbf{s}_{y_k} \rvert^2}  } \geq \gamma,
\label{stpUL}
\end{align}
\noindent while for the DL signal received at the STA, using (\ref{snrdl})
\begin{align}
& \mathbb{P}\left(\mbox{SINR}_{y_j} \geq \gamma \right) = \nonumber \\
& \frac{\xi_{x_i y_j}\ell \left( x_i, y_j \right) \lvert \mathbf{W}^{}_{y_j} \mathbf{H}_{x_i y_j}\mathbf{V}^{}_{x_i}\textbf{s}_{x_i} \rvert^2 }{ \norm{\mathbf{n}_{y_j}}^2 + {\ell\left( y_j, y_j \right) \lvert \mathbf{W}^{}_{y_j} \mathbf{H}_{y_j y_j}\mathbf{V}^{}_{y_j}\textbf{s}_{y_j}\rvert^2} + {\sum\limits_{x_k \in \tilde{\Phi}_a, i \neq k }^{  } \ell \left( x_k, y_j \right) \lvert \mathbf{W}^{}_{y_j} \mathbf{H}_{x_k y_j} \mathbf{V}^{}_{x_k} \textbf{s}_{x_k} \rvert^2} } \geq \gamma,
\label{stpDL}
\end{align}
\noindent given the uniqueness of the interference pattern (due to the independent set of interference sources) and the SI in the UL and the DL, a FD transmission is successful with probability $\mathbb{P}_{FD}$ according to

\begin{lemma}
	\label{stplem}
	The successful transmission probability of a bidirectional FD transmission is
	\begin{align}
	\mathbb{P}_{FD} & = \mathbb{P}\left(\mbox{SINR}_{{x_i}} \geq \gamma \right) \mathbb{P}\left(\mbox{SINR}_{{y_j}} \geq \gamma \right) \nonumber\\
	& = \exp\left(- 2 \frac{\gamma \norm{\mathbf{n}_{x_i} }^2}{ \xi_{x_i y_j} \ell \left( y_j, x_i \right) } - \frac{1}{\pi} \left[\ln\left(1 + \frac{\gamma \left( \frac{1}{ \tilde{\lambda}_s \pi } \right)^{-\alpha} }{\xi_{x_i y_j} \ell \left( y_j, x_i \right)}\right) - \ln\left(1 + \frac{\gamma \left( \frac{1}{ \tilde{\lambda}_a \pi } \right)^{-\alpha} }{\xi_{x_i y_j} \ell \left( y_j, x_i \right)}\right)\right] \right) \nonumber\\
	& \exp\left(- 2 \left( \frac{1}{ \tilde{\lambda}_s \pi } + \frac{1}{ \tilde{\lambda}_a \pi } \right) + 2 \left[ \left(\frac{\gamma \left(\frac{1}{ \tilde{\lambda}_s \pi }\right)^{1-\alpha} }{ \xi_{x_i y_j}\ell \left( y_j, x_i \right) }\right)^2 \arctan \frac{ \frac{1}{ \tilde{\lambda}_s \pi } }{ \left(\frac{\gamma \left(\frac{1}{ \tilde{\lambda}_s \pi }\right)^{1-\alpha} }{\xi_{x_i y_j} \ell \left( y_j, x_i \right) }\right)^2 }  \right. \right. \nonumber \\
	 & \left. \left.  + \left(\frac{\gamma \left(\frac{1}{ \tilde{\lambda}_a \pi }\right)^{1-\alpha} }{\xi_{x_i y_j} \ell \left( y_j, x_i \right) }\right)^2 \arctan \frac{ \frac{1}{ \tilde{\lambda}_a \pi } }{ \left(\frac{\gamma \left(\frac{1}{ \tilde{\lambda}_a \pi }\right)^{1-\alpha} }{ \xi_{x_i y_j} \ell \left( y_j, x_i \right) }\right)^2 } \right]\right).
	\label{jointprob}
	\end{align}
\end{lemma}
\normalsize
\noindent \textit{Proof}. Since the UL and DL channels are independent, the STP in the UL is also independent of the STP in the DL. Therefore, we proceed by first computing the STP in Eqn.~(\ref{stpUL})  for the case of UL transmission. From Eqn.~(\ref{stpUL}), let $P^t = 1$, $\Psi_{x_i, x_i} = \mathbb{E} \lvert \mathbf{W}^{}_{x_i} \mathbf{H}_{x_i x_i}\mathbf{V}^{}_{x_i}\textbf{s}_{x_i} \rvert^2 $, and $ \Psi_{y_k, x_i} = \mathbb{E} \lvert \mathbf{W}^{}_{x_i} \mathbf{H}_{y_k x_i} \mathbf{V}^{}_{y_k} \textbf{s}_{y_k} \rvert^2 $, $\mathbb{P}\left(\mbox{SINR}_{{{x_i}}} \geq \gamma \right)$ is written as:
\small
\begin{align}
\label{stpFD}
& \mathbb{E} \lvert \mathbf{H}_{y_j, x_i}\mathbf{V}^{}_{y_j}\textbf{s}_{y_j} \mathbf{W}^{}_{x_i}\rvert^2 \geq \frac{\gamma}{  {\xi_{x_i y_j}} \ell \left( y_j, x_i \right) }\left(  \norm{\mathbf{n}_{x_i} }^2 + \ell\left( x_i, x_i \right)\Psi_{x_i, x_i} + \sum\limits_{y_k \in \tilde{\Phi}_s, j \neq k }^{  } \ell \left( y_k, x_i \right) \Psi_{y_k, x_i}  \right) \\
& \myeq \exp\left(- \frac{\gamma \norm{\mathbf{n}_{x_i} }^2}{{\xi_{x_i y_j}} \ell \left( y_j, x_i \right) } \right) \mathbb{E}\left[\exp\left( - \frac{\gamma \ell\left( x_i, x_i \right)}{ {\xi_{x_i y_j}} \ell \left( y_j, x_i \right) } \Psi_{x_i, x_i} \right) \right] \mathbb{E}\left[ \exp\left(- \sum\limits_{y_k \in \tilde{\Phi}_s, j \neq k }^{  } \frac{\gamma \ell \left( y_k, x_i \right)}{ {\xi_{x_i y_j}}\ell \left( y_j, x_i \right) }\Psi_{y_k, x_i} \right) \right]  \nonumber \\
& \beq \exp\left(- \frac{\gamma \norm{\mathbf{n}_{x_i} }^2}{ {\xi_{x_i y_j}} \ell \left( y_j, x_i \right) } \right) \exp\left( - \frac{\gamma \ell\left( x_i, x_i \right)}{ {\xi_{x_i y_j}} \ell \left( y_j, x_i \right) } \Psi_{x_i, x_i} \right) \exp\left(- \mathbb{E}_{\tilde{\Phi}_s} \left[\sum\limits_{y_k \in \tilde{\Phi}_n, j \neq k }^{  } \frac{\gamma \ell \left( y_k, x_i \right)}{ {\xi_{x_i y_j}} \ell \left( y_j, x_i \right) } \Psi_{y_k, x_i}\right] \right) \nonumber \\
& \ceq \exp\left(- \frac{\gamma \norm{\mathbf{n}_{x_i} }^2}{ {\xi_{x_i y_j}} \ell \left( y_j, x_i \right) } \right)\exp\left( - \frac{\gamma \ell\left( x_i, x_i \right)}{ {\xi_{x_i y_j}} \ell \left( y_j, x_i \right) } \kappa\rho \right) \exp\left( - \mathbb{E}_{\tilde{\Phi}_s} \left[ \prod_{ y_k \in \tilde{\Phi}_s, j \neq k }^{ } \frac{1}{1 + \frac{\gamma \ell \left( y_k, x_i \right)}{ {\xi_{x_i y_j}} \ell \left( y_j, x_i \right) } }\right] \right) \nonumber\\
& \deq \exp\left(- \frac{\gamma \norm{\mathbf{n}_{x_i} }^2}{ {\xi_{x_i y_j}} \ell \left( y_j, x_i \right) } \right) \exp\left( - \frac{\gamma \ell\left( x_i, x_i \right)}{ {\xi_{x_i y_j}} \ell \left( y_j, x_i \right) } \kappa\rho \right) \exp\left( - \mathbb{E}_{\tilde{\Phi}_s} \left[ \sum\limits_{y_k \in \tilde{\Phi}_s, j \neq k }^{  } \ln\left( 1 + \frac{\gamma \ell \left( y_k, x_i \right)}{ {\xi_{x_i y_j}} \ell \left( y_j, x_i \right) } \right)\right] \right) \nonumber \\
& \eeq \exp\left(- \frac{\gamma \norm{\mathbf{n}_{x_i} }^2}{ {\xi_{x_i y_j}} \ell \left( y_j, x_i \right) } \right) \exp\left( - \frac{\gamma \ell\left( x_i, x_i \right)}{ {\xi_{x_i y_j}} \ell \left( y_j, x_i \right) } \kappa\rho \right) \exp\left( - \tilde{\lambda}_s \int_{ {\mathbb{R}}^d }^{ } \ln\left( 1 + \frac{\gamma \lVert  y_k - x_i \rVert^{-\alpha}}{ {\xi_{x_i y_j}} \ell \left( y_j, x_i \right) } \right) \mbox{ d}\lVert  y_k - x_i \rVert \right) \nonumber\\
& \feq \exp\left(- \frac{\gamma \norm{\mathbf{n}_{x_i} }^2}{ {\xi_{x_i y_j}} \ell \left( y_j, x_i \right) } \right) \exp\left( - \frac{\gamma \ell\left( x_i, x_i \right)}{ {\xi_{x_i y_j}} \ell \left( y_j, x_i \right) } \kappa\rho \right) \exp\left(- \tilde{\lambda}_s\lVert  y_k - x_i \rVert  \ln\left(1 + \frac{\gamma \lVert  y_k - x_i \rVert^{-\alpha}}{ {\xi_{x_i y_j}} \ell \left( y_j, x_i \right) } \right)  \right. \nonumber \\ & \left. - 2\lVert  y_k - x_i \rVert + 2 \left( \frac{\gamma \lVert  y_k - x_i \rVert^{1 -\alpha}}{ {\xi_{x_i y_j}}\ell \left( y_j, x_i \right) }\right)^2 \arctan \frac{\lVert  y_k - x_i \rVert}{\left( \frac{\gamma \lVert  y_k - x_i \rVert^{1 -\alpha}}{ {\xi_{x_i y_j}}\ell \left( y_j, x_i \right) }\right)^2}\right),  \nonumber
\end{align}
\normalsize
\noindent where (a) follows from the exponential distribution of the signal power $\lvert \mathbf{W}^{}_{x_i} \mathbf{H}_{y_j, x_i}\mathbf{V}^{}_{y_j}\textbf{s}_{y_j} \rvert^2$, (b) holds by taking the expectation with respect to the SI power $\Psi_{x_i, x_i}$ and the interference power $\Psi_{y_k, x_i}$,  (c)-(d) follows from taking the expectation of the interference terms over the PPP $\tilde{\Phi}_n$ of active transmitters, (e) is obtained by applying Campbell's theorem $\mathbb{E} \left[ \sum_{x \in \Phi}^{} f\left(x\right) \right] = \lambda \int_{\mathbb{R}^2}^{} f\left(x\right) dx $ \cite{stoyan} and (f) follows from the integral transformation of \cite[Eqn. 2.733.1]{tableInt}. The proof of the DL STP $\mathbb{P}\left(\mbox{SINR}_{y_j} \geq \gamma \right)$ follows similar steps. By combining $\mathbb{P}\left(\mbox{SINR}_{{ {x_i}}} \geq \gamma \right)$ and $\mathbb{P}\left(\mbox{SINR}_{y_j} \geq \gamma \right)$, Eqn.~(\ref{jointprob}) is obtained where it is assumed that the target SINR $\gamma$ and noise are identical for both UL and DL, causing the SI at both UL and DL to cancel out. \qed

\subsection{Spatial Density of Throughput (SDT)}{\label{fdthruput}}

Spatial density of throughput (SDT) measures the average throughput performance per unit area \cite{yingzhel}, \cite{tnguyen}. In other words, it captures the tradeoff between channel access probabilities (CAPs) $p_{y_j}$ and $p_{x_i}$ defined in Eqs.~(\ref{probofy}) and (\ref{activeap}), respectively, and STP. When more nodes have high CAP to transmit, interference increases and the probability (or STP) of achieving SINR threshold $\gamma$ decreases. On the other hand, when fewer users transmit concurrently, interference is less of a concern and that implies high SINR. To capture this tradeoff, the SDT for a HD transmission mode in the DL, the throughput per unit area is
\begin{equation}
\Upsilon^{DL}_{HD} = \tilde{\lambda}_{a} \times \log(1 + \gamma) \times \mathbb{P}\left(\mbox{SINR}_{{y_j}} \geq \gamma \right) \qquad \mbox{nats/sec/Hz},
\end{equation}
\noindent where $\tilde{\lambda}_{a}$ is the density of concurrently transmitting APs according to Equation~(\ref{activeap}), $\mathbb{P}\left(\mbox{SINR}_{{y_j}} \geq \gamma \right)$ is derived in Lemma~\ref{stplem}. Similarly, for UL HD transmission, we have 
\begin{equation}
\Upsilon^{UL}_{HD} = \tilde{\lambda}_{s} \times \log(1 + \gamma) \times \mathbb{P}\left(\mbox{SINR}_{{x_i}} \geq \gamma \right) \qquad \mbox{nats/sec/Hz}.
\end{equation}

\noindent Given the probability of successful transmission $\mathbb{P}_{FD}$ of a FD transmission, the spatial density of network throughput of a FD transmission can be written as
\begin{equation}
\Upsilon_{FD}  = \tilde{\lambda}_{FD} \log(1 + \gamma)\mathbb{P}_{FD} \qquad \qquad \qquad \qquad \mbox{nats/sec/Hz}, 
\label{fdThruput}
\end{equation}
\noindent where $\tilde{\lambda}_{\mbox{FD}} $ is given in Eqn.~(\ref{fddensity}).

\section{Joint User-AP Association and PCS Threshold Framework}{\label{framework}}
In this section, we address the user-AP association problem with the objective of maximizing the throughput per network area given in Eqn.~(\ref{fdThruput}) for FD networks. The overall goal here is to find both an optimal association and a PCS threshold $\Gamma$ that maximizes the average throughput in the entire network. Generally, user-AP association is a NP-hard combinatorial problem. Herein, we seek a solution via Lagrangian duality theory by relaxing the binary association variable. Once a solution is realized for the user-AP association, the PCS threshold is then optimized. However, the overall joint solution may not be optimal. In a FD WLAN, if STA $y_j$ associates with AP $x_i$, the mean throughput per unit area of the FD transmissions is given by Eqn.~(\ref{fdThruput}). To improve average performance by optimally selecting an AP, the optimization problem is formulated as:
\begin{subequations}
	\label{userAssocOptimizationSP}
	\begin{align}
	& \text{maximize} & & \Upsilon_{FD},  \label{objectveFuncSP} \\
	& \text{subject to} & & \sum\limits_{x_i \in {\Phi}_a } \xi_{x_i y_j} = 1. \qquad \forall y_j \in \Phi_{n} \label{sumassoc}\\
	& & & \xi_{x_i y_j} \in \{0, 1\} \label{assocConstraintSP} \\
	& & & \Gamma \leq \xi_{x_i y_j}\lVert  x_i - y_j \rVert^{-\alpha} \frac{1}{ \left(1 + P^t \gamma^{\frac{1}{\alpha}}\right)^{\alpha} } \qquad \forall y_j \in \Phi_{a}, \forall x_i \in \Phi_{n}.
	\end{align}
\end{subequations}
\noindent The objective in Eqn.~(\ref{objectveFuncSP}) indicates that when an STA $y_j$ associates with AP $x_i$, the expected rate is $\Upsilon_{FD}$. The solution to this problem is realized by finding the optimal association ${\xi}^*_{x_i y_j}$ and optimal PCS threshold $\Gamma$ that maximize the spatial density of FD throughput $\Upsilon_{FD}$. To obtain a solution that jointly assigns STAs to the APs and design the PCS threshold such that the average performance is maximized, the original problem in (\ref{userAssocOptimizationSP}) is decomposed into two coupled subproblems addressing the user-AP association and the PCS threshold selection as
\begin{subequations}
	\label{userAssoc}
	\begin{align}
	& \max\limits_{\xi_{x_i y_j}} & & \tilde{\lambda}_{FD} \log(1 + \gamma)\mathbb{P}_{FD},  \label{objectveFuncSP2} \\
	& \text{subject to} & & \sum\limits_{x_i \in {\Phi}_a } \xi_{x_i y_j} = 1. \qquad \forall y_j \in \Phi_{n} \label{sumassoc2}\\
	& & & \xi_{x_i y_j} \in \{0, 1\} \label{assocConstraintSP2} \\
	& & & \Gamma \leq \xi_{x_i y_j}\lVert  x_i - y_j \rVert^{-\alpha} \frac{1}{ \left(1 + P^t \gamma^{\frac{1}{\alpha}}\right)^{\alpha} } \qquad \forall y_j \in \Phi_{a}, \forall x_i \in \Phi_{n},
	\end{align}
\end{subequations}
\noindent and
\begin{subequations}
	\label{pcsselection}
	\begin{align}
	& \max\limits_{\Gamma}  & & \tilde{\lambda}_{FD} \log(1 + \gamma)\mathbb{P}_{FD},  \label{objectveFuncSP3} \\
	& \text{subject to} & & \Gamma \leq \xi_{x_i y_j}\lVert  x_i - y_j \rVert^{-\alpha} \frac{1}{ \left(1 + P^t \gamma^{\frac{1}{\alpha}}\right)^{\alpha} } \qquad \forall y_j \in \Phi_{a}, \forall x_i \in \Phi_{n},
	\end{align}
\end{subequations}

\noindent respectively. Subsequently, solutions to the above subproblems are sought for respectively in the following subsections. The overall goal of decomposing the problem into subproblems is to first obtain a solution for the optimal user-AP association under a globally fixed PCS threshold $\Gamma$. Then, given the optimal association factor $\xi^*_{x_i y_j}$, the PCS threshold is optimized.

\subsection{User Association Problem}{\label{apsol}}

To solve problem (\ref{userAssoc}), we can relax constraint $ \xi_{x_i y_j} \in \{0, 1\}$ in (\ref{assocConstraintSP2}) from a binary value to take on continuous values between 0 and 1, primarily due to the complexity of solving this type of combinatorial problem. Therefore, by setting $\lVert  x_i - y_j \rVert = \frac{1}{{\lambda}_{FD} \pi}$ based on Lemma~\ref{ssflemma}, Problem~(\ref{userAssoc}) is reformulated as 
\begin{subequations}
	\label{userAssoc2}
	\begin{align}
	& \max\limits_{0 \leq \xi_{x_i y_j} \leq 1} & & \tilde{\lambda}_{FD} \log(1 + \gamma)\mathbb{P}_{FD},  \label{objectveFuncSP21} \\
	& \text{subject to} & & \sum\limits_{x_i \in {\Phi}_a } \xi_{x_i y_j} = 1. \qquad \forall y_j \in \Phi_{n} \label{sumassoc21}\\
	& & & \Gamma \left( \frac{1 + P^t \gamma^{\frac{1}{\alpha}}}{{\lambda}_{FD} \pi} \right)^{\alpha}  \leq \xi_{x_i y_j} \qquad \forall y_j \in \Phi_{a}, \forall x_i \in \Phi_{n},
	\label{pcsthrcon}
	\end{align}
\end{subequations}
\noindent and using Lagrangian dual technique \cite{qyerong}, \cite{yinjunL} a solution is feasible. The solution is given by
\begin{theo}
	The optimal user-AP association policy that maximizes FD throughput $\Upsilon_{FD}$ is
	\small
	\begin{align}
	& {\xi}^*_{x_i y_j} \nonumber \\ 
	& = \underset{y_j}{\arg\max}\mbox{ } \left\{ \tilde{\lambda}_{FD} \log(1 + \gamma)\mathbb{P}_{FD} +  \left(\sum\limits_{x_i \in {\Phi}_a } \xi_{x_i y_j} - 1\right)\delta + \left( \Gamma \left( \frac{1 + P^t \gamma^{\frac{1}{\alpha}}}{{\lambda}_{FD} \pi} \right)^{\alpha}  - \xi_{x_i y_j}\right) \eta \right\}.  
	\label{optimalindicator}
	\end{align}
	\label{theo1}
	\normalsize
\end{theo}
\noindent \textit{Proof}. Let $\delta$ and $\eta$ be the Lagrangian multipliers associated with constraints (\ref{sumassoc21}) and (\ref{pcsthrcon}), respectively, the Lagrangian dual of problem (\ref{userAssocOptimizationSP}) is
\begin{equation}
\mathcal{L}\left(\xi_{x_i y_j}, \delta, \eta \right) = \tilde{\lambda}_{FD} \log(1 + \gamma)\mathbb{P}_{FD} +  \left(\sum\limits_{x_i \in {\Phi}_a } \xi_{x_i y_j} - 1\right)\delta + \left( \Gamma \left( \frac{1 + P^t \gamma^{\frac{1}{\alpha}}}{{\lambda}_{FD} \pi} \right)^{\alpha}  - \xi_{x_i y_j}\right) \eta,
\label{lagassoc}
\end{equation}
\noindent and the dual objective becomes
\begin{equation}
g\left(\delta, \eta\right) = \underset{\xi_{x_i y_j}}{\max}\mbox{ } \mathcal{L}\left(\xi_{x_i y_j}, \delta, \eta \right),
\end{equation}
\noindent which results to the dual optimization problem
\begin{align}
& \text{minimize} & &  g\left(\delta, \eta\right) \nonumber\\
& \text{subject to} & &  \delta, \eta \geq 0.
\end{align}

The user association is obtained by iterating the \textit{necessary conditions} until the rate utility (\ref{objectveFuncSP21}) stops improving. The optimal value of $\delta$ and $\eta$ can be obtained via \textit{subgradient method} \cite{yinjunL} since the Lagrangian function of the dual problem is non-differentiable. Given a dynamic step size $\phi (k)$, the Lagrangian multipliers are updated as:

\begin{equation}
\delta^{k + 1}_{} = \left[ \delta^k_{} - \phi(k) \left(  \sum\limits_{x_i \in {\Phi}_a } \delta^{k + 1}_{} - 1  \right) \right]^+,
\label{piupdate}
\end{equation}
\noindent and
\begin{equation}
\eta^{k + 1}_{} = \left[ \eta^k_{} - \phi(k) \left( \Gamma \left( \frac{1 + P^k \gamma^{\frac{1}{\alpha}}}{{\lambda}_{FD} \pi} \right)^{\alpha}  - \xi_{x_i y_j}\right)  \right]^+,
\label{etaupdate}
\end{equation}
\normalsize
\noindent the step size $\phi(k)$ is updated at each iteration. The above solution is suboptimal as a result of decoupling threshold selection from the original problem and by relaxing constraint (\ref{assocConstraintSP2}). \qed

\subsection{Optimal PCS Threshold Selection}{\label{pcsthsol}}

With the AP association solution ${\xi}^*_{x_i y_j}$ obtained in \textbf{Theorem~\ref{theo1}} under Section~(\ref{apsol}), the next objective is to solve the PCS threshold selection problem in  (\ref{pcsselection}). Since ${\xi}^*_{x_i y_j}$ is obtained, the PCS threshold selection subproblem (\ref{pcsselection}) is reformulated as
\begin{subequations}
	\label{pcsselection2}
	\begin{align}
	& \max\limits_{\Gamma}  & & \Upsilon(\Gamma) = \tilde{\lambda}_{FD} \log(1 + \gamma)\mathbb{P}_{FD},  \label{objectveFuncSP32} \\
	& \text{subject to} & & \Gamma \leq {\xi}^*_{x_i y_j} \left( \frac{1 + P^t \gamma^{\frac{1}{\alpha}}}{{\lambda}_{FD} \pi} \right)^{-\alpha} \qquad \forall y_j \in \Phi_{a}, \forall x_i \in \Phi_{n}.
	\label{pccon}
	\end{align}
\end{subequations}
\noindent Let $\Upsilon(\Gamma)$ denote the objective function in (\ref{objectveFuncSP32}). Since the density of active FD nodes $\tilde{\lambda}_{FD}$ obtained in \textbf{Lemma~\ref{denFD}} is an increasing function of $\Gamma$, we obtain the first-order and the second-order partial derivatives of $\Upsilon(\Gamma)$ with respect to $\Gamma$. Since the objective function $\Upsilon(\Gamma)$ is twice differentiable and $\frac{\partial^2 \Upsilon(\Gamma)}{\partial \Gamma^2}$ is continuous in $\Gamma^*$, the solution to the PCS threshold selection objective $\Upsilon(\Gamma)$ is numerically obtained using the truncated Newton Method (also known as \textit{line search conjugate gradient} method)  \cite{numoptz} with incremental Newton search direction \cite{yanlin}, \cite{kshen}, which leads to the search iteration policy:
\begin{equation}
\label{iterpolicy}
\Gamma^{\left(k + 1\right)} = \Gamma^{\left(k\right)} + \epsilon_k \underbrace{\frac{\partial \Upsilon(\Gamma)}{\partial \Gamma} \bigg/ \norm{\frac{\partial^2 \Upsilon(\Gamma)}{\partial \Gamma^2}}}_{\varpi},
\end{equation}

\noindent where $\epsilon_k$ is the step length and $\varpi$ is the Newton ascent search direction. The step length is chosen through the well-known \textit{backtracking} approach \cite{yanlin}, \cite{numoptz}. To terminate the Newton iteration at an approximate (or inexact) solution \cite{numoptz}, we define the termination criterion
\begin{equation}
\norm{ \frac{\partial^2 \Upsilon(\Gamma_k)}{\partial \Gamma^2_k} \varpi + \frac{\partial \Upsilon(\Gamma_k)}{\partial \Gamma_k} } \leq \nu_k \norm{\frac{\partial \Upsilon(\Gamma_k)}{\partial \Gamma_k}},
\label{termcri}
\end{equation}

\noindent where $\nu_k, 0 \leq \nu_k < 1$ is the \textit{forcing sequence}, which can be chosen to achieve ``superlinear'' convergence rate as thus\cite{numoptz}:
\begin{equation}
\label{fors}
\nu_k =  \min\left(0.5, \sqrt{\norm{\frac{\partial \Upsilon(\Gamma_k)}{\partial \Gamma_k}}}\right).
\end{equation}

\noindent Finally, since $\Gamma$ is bounded by ${\xi}^*_{x_i y_j} \left( \frac{1 + P^t \gamma^{\frac{1}{\alpha}}}{{\lambda}_{FD} \pi} \right)^{-\alpha}$ according to constraint (\ref{pccon}), the solution obtained from the above Newton iterative method is verified against constraint (\ref{pccon}). Therefore, the PCS threshold selection step is terminated when either termination criterion in Eqn.~(\ref{termcri})  or the following necessary condition is satisfied:
\begin{equation}
\Gamma_k - {\xi}^*_{x_i y_j} \left( \frac{1 + P^t \gamma^{\frac{1}{\alpha}}}{{\lambda}_{FD} \pi} \right)^{-\alpha} = 0,
\label{concri}
\end{equation} 
\noindent which is introduced as an additional criterion without loss of generality, to ensure that $\Gamma_k$ satisfies the constraint. In general, a solution based on Newton's method may not necessarily converge.

\subsection{Joint User-AP Association and PCS Threshold Selection Algorithm}

The proposed algorithm to jointly solve user-AP association and PCS threshold selection is presented in Algorithm~\ref{APassocwithpcs}. First the user-AP association problem is solved iteratively to obtain ${\xi}^*_{x_i y_j}$, and once ${\xi}^*_{x_i y_j}$ is determined, the PCS threshold $\Gamma$ selection problem is solved using the Newton iteration method in Eqn.~(\ref{iterpolicy}).

\begin{rem}
	In wireless networks, user association with the APs takes place before PCS threshold selection. Performing the user-AP association first is to ensure that users are distributed among best serving APs. Then, by further optimizing the PCS threshold, interference from concurrent transmitters is reduced because the PCS threshold determines the degree of spatial reuse and the number of concurrent transmitters per time-slot.
\end{rem}

\begin{figure}[!h]
	\begin{algorithm}[H]
		\caption{Joint User-AP Association and PCS Threshold Optimization (JAPO)}
		\label{APassocwithpcs}
		\textbf{Initialize} $\Gamma$, $k = 0$, $\phi\left(k\right)$, $\eta$, and $\delta$\\
		For fixed $\Gamma$, obtain association variable ${\xi}^*_{x_i y_j}$:\\
		\Repeat{${\xi}^*_{x_i y_j}$ $\text{converges}$}{
			Calculate ${\xi}^*_{x_i y_j}$ using (\ref{optimalindicator})\\
			Update $\delta\left(k+1\right)$ using (\ref{piupdate}) \\
			Update $\eta\left(k + 1\right)$ using (\ref{etaupdate}) \\
			$k \leftarrow k + 1$
		}
		For a given ${\xi}^*_{x_i y_j}$ solve for $\Gamma^*$:\\
		Set $k = 0$, Calculate $\varpi$\\
		\Repeat{ (\ref{termcri}) or (\ref{concri}) is satisfied }{ 
			Compute Eqn.~(\ref{iterpolicy})\\
			Update $\Gamma_k$ and $k = k + 1$
		}
	\end{algorithm}
\end{figure}

\subsection{User-AP Association under Strongest Signal First (SSF)}{\label{analysis}}

For comparison purposes, we consider the method currently used in WLAN\cite{weili} where users select the closest AP that offers strongest received signal strength (RSS) and the PCS threshold is fixed in the network. Given the path loss model in Eqn.~(\ref{pathloss}), it is apparent that selecting an AP based on the SSF (or strongest RSS) means that an STA selects the closest AP. Let each user-AP pair be at distance $\lVert  y_j - x_i \rVert$ (i.e., distance between one STA and one AP) in the network. From Lemma~\ref{ssflemma}, it is established that $\lVert  y_j - x_i \rVert$ has a probability distribution characterized as
\begin{equation}
f\left( \lVert  y_j - x_i \rVert \right) = \frac{2\lambda_{\mbox{FD}}\pi\lVert  y_j - x_i \rVert^2}{\lVert  y_j - x_i \rVert} \exp\left({-\lambda_{\mbox{FD}}\pi\lVert  y_j - x_i \rVert^2}\right)
\label{ssfassociation}
\end{equation}


\noindent where $\lambda_{\mbox{FD}}$ represents the density of FD locations in the network, which is obtained through \textit{superposition} of the two independent node densities $\lambda_s$ and $\lambda_a$ in Eqn.~(\ref{fddens}). Since under the SSF association scheme, an STA $y_j$ forms a FD pair with the closest AP $x_i$, Eqn.~(\ref{ssfassociation}) is the distribution of SSF association in the network.

Consequently, for an STA at point $y_j$ associated with an AP at point $x_i$ according to Lemma~\ref{ssflemma}, the spatial average of the FD rate is immediate from

\begin{theo}
	\label{ssfmeanrate}
	The achievable spatial mean rate of FD links under SSF association is
	\begin{align}
	& \Lambda^{ssf}_{FD}  = \int_{\lVert  y_j - x_i \rVert = 0}^{ \infty } \Upsilon_{FD}  \mbox{ d} \lVert  y_j - x_i \rVert = \tilde{\lambda}_{\mbox{FD}}\log\left(1 + \gamma\right)\nonumber\\
	&  \int_{\lVert  y_j - x_i \rVert = 0}^{ \infty } \exp\left(- 2 \frac{\gamma \norm{\mathbf{n}_{x_i} }^2}{ \ell \left( y_j, x_i \right) } - \frac{1}{\pi} \left[\ln\left(1 + \frac{\gamma \left( \frac{1}{ \tilde{\lambda}_s \pi } \right)^{-\alpha} }{\ell \left( y_j, x_i \right)}\right) - \ln\left(1 + \frac{\gamma \left( \frac{1}{ \tilde{\lambda}_a \pi } \right)^{-\alpha} }{\ell \left( y_j, x_i \right)}\right)\right]\right) \nonumber\\
	& e^{\left(- 2 \left( \frac{1}{ \tilde{\lambda}_s \pi } + \frac{1}{ \tilde{\lambda}_a \pi } \right) + 2 \left[ \left(\frac{\gamma \left(\frac{1}{ \tilde{\lambda}_s \pi }\right)^{1-\alpha} }{ \ell \left( y_j, x_i \right) }\right)^2 \arctan \frac{ \frac{1}{ \tilde{\lambda}_s \pi } }{ \left(\frac{\gamma \left(\frac{1}{ \tilde{\lambda}_s \pi }\right)^{1-\alpha} }{ \ell \left( y_j, x_i \right) }\right)^2 }  \right. \right.
		\left. \left.  + \left(\frac{\gamma \left(\frac{1}{ \tilde{\lambda}_a \pi }\right)^{1-\alpha} }{ \ell \left( y_j, x_i \right) }\right)^2 \arctan \frac{ \frac{1}{ \tilde{\lambda}_a \pi } }{ \left(\frac{\gamma \left(\frac{1}{ \tilde{\lambda}_a \pi }\right)^{1-\alpha} }{ \ell \left( y_j, x_i \right) }\right)^2 } \right]\right)} \mbox{ d} \lVert  y_j - x_i \rVert
	\label{ssfmeanrat}
	\end{align}
\end{theo}
\normalsize
\noindent \textit{Proof}. Setting ${\xi_{x_i y_j}} = 1$ and substituting Eqn.~(\ref{jointprob}) from Lemma~\ref{stplem} in Eqn.~(\ref{fdThruput}) and integrating the mean rate utility over the SSF association distribution in Eqn.~(\ref{ssfassociation}), Eqn.~(\ref{ssfmeanrat}) is obtained.\qed

\begin{rem}
	While Eqn.~(\ref{ssfmeanrat}) is not solvable in closed form, it can be solved numerically with $\ell \left( y_j, x_i \right) = \lVert  y_j - x_i \rVert^{-\alpha}$ . 
\end{rem}

\section{Numerical Analysis and Observations}\label{chjperformanceI}

\subsection{System Setup and Parameters}

For simulation purposes, we consider a 2D wireless network where AP and STA locations are generated as realizations of independent PPPs denoted as $\lambda_s$ and $\lambda_a$, respectively. Simulation is performed for various STA densities $\lambda_s$ while the density of APs is fixed at $\lambda_a = 0.3$. The path loss exponent $\alpha = 3.4$, and the noise variance $\sigma^2 = -100$ dBm throughout the simulation. For the fixed PCS threshold case, $\Gamma = -70$ dBm and the CSR is computed based on $\Gamma$. The transmit power $P_t$ of APs and the STAs is fixed as $100$mW (20 dBm) and APs and STAs are equipped with $M  = 4$ and $N = 2$ antennas, respectively. The SINR threshold $\gamma$ is assumed identical for both the UL and the DL transmissions, and it is chosen for specific WLAN transmission rates (see \cite[Table II]{weili}). For the FD self-interference (SI) power, the shape parameter $\kappa$ and the scale parameter $\rho$ are computed according to Eqn.~(\ref{siparams}) with mean $\mu$ and variance $\psi^2$ obtained from Eqn.~(\ref{meanvarsi}) for Ricean $K$-factor $K = 1$ \cite{mduarte} and SI attenuation factor $\Omega = -80$ dB \cite{atzeni}. 

\begin{figure}[!h]
	\centering
	\includegraphics[width=5in, height=3in]{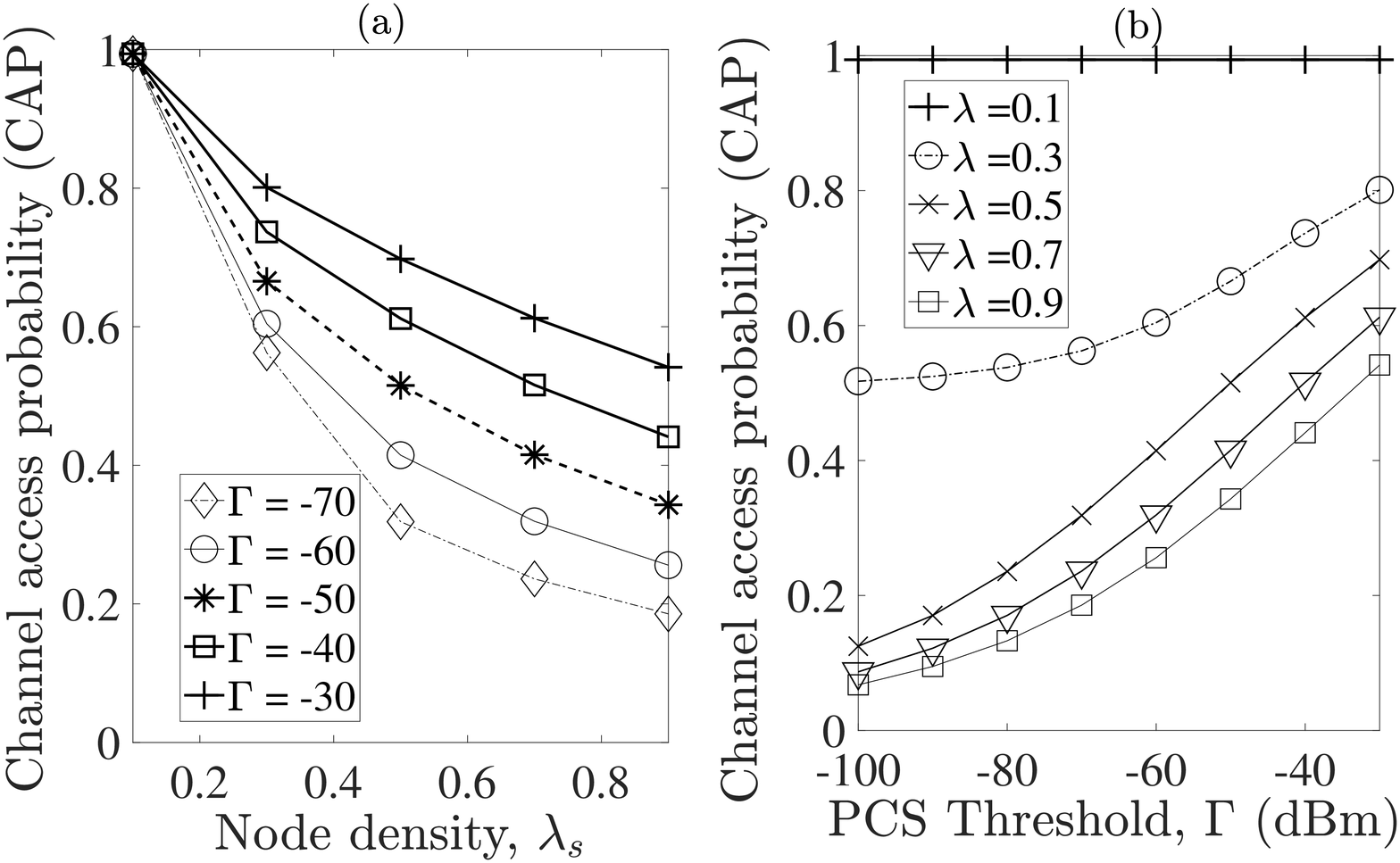}
	\caption{Channel access probability, $p_{_yi}$ versus (a) Node density $\lambda_s$ versus PCS threshold $\Gamma$ and (b) PCS threshold $\Gamma$ as a function of node density $\lambda_s$.}
	\label{fig_cap}
\end{figure}

\subsection{Validation, Performance Gains and Discussion}

For each node density $\lambda_s$, the simulation results are averaged over $10^4$ network realizations. To evaluate the performance of the proposed joint AP association and PCS threshold selection algorithm (JAPO) in Algorithm~\ref{APassocwithpcs}, its performance is compared to the strongest signal first (SSF) scheme, which is the default AP association scheme in current WLAN systems \cite{weili} and analyzed in Theorem~\ref{ssfmeanrate}. The second scheme considered is the case of optimizing the AP association according Theorem~\ref{theo1} without PCS threshold optimization and it termed ``FD Assoc. with fixed PCS threshold (FD Assoc. w/fixed PCS)." The other scenario considered is the half-fuplex case of the proposed JAPO algorithm. The performance metric of interest according to the objective in Eqn.~(\ref{objectveFuncSP}), is the spatial average throughput measured in nats/sec/Hz.

\begin{figure}[!h]
	\centering
	\includegraphics[width= 5in, height=3in]{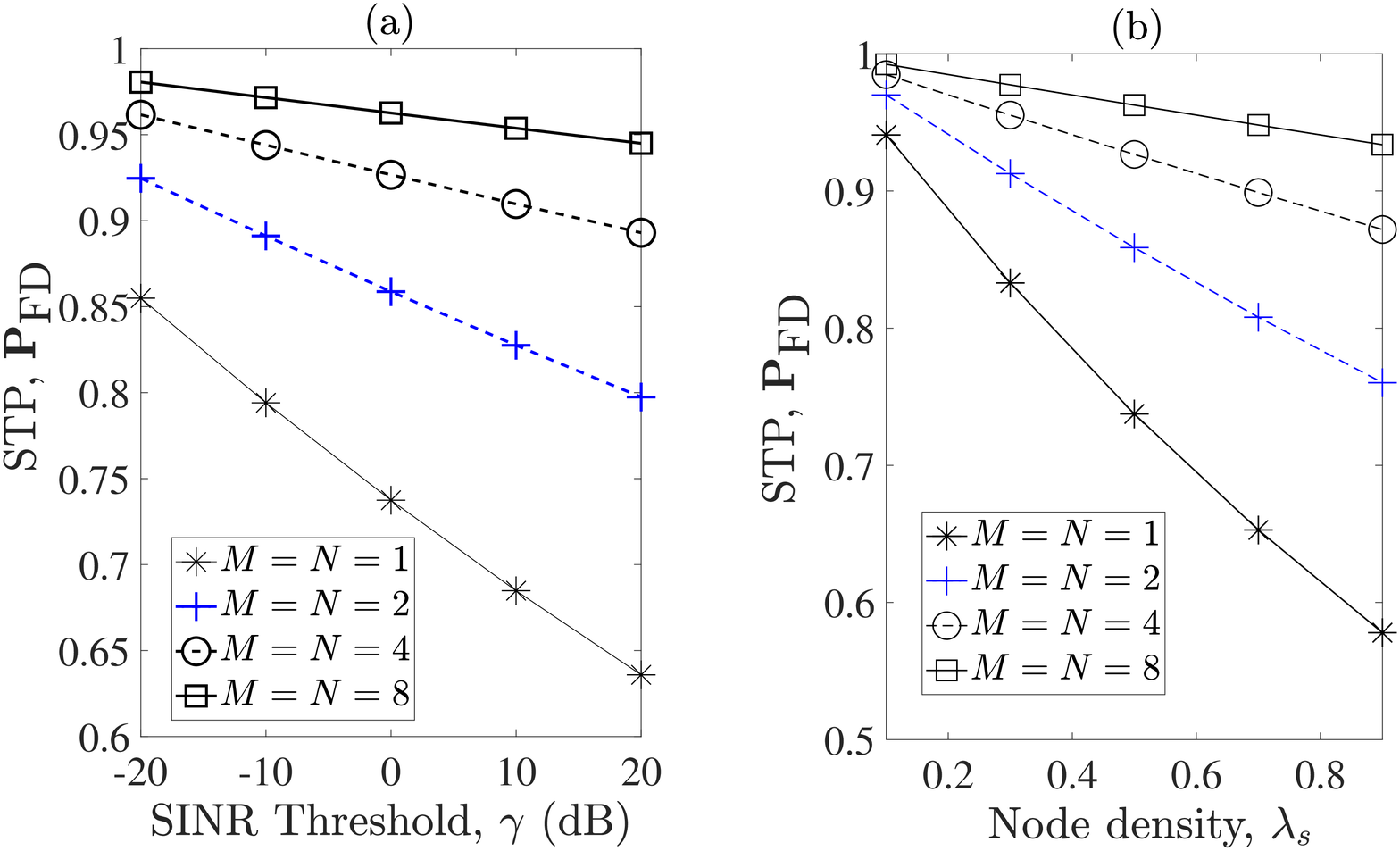}
	\caption{(a) Successful transmission probability (STP) versus SINR values at node density $\lambda_s = 0.5$ (b) STP versus node density $\lambda_s$, SINR $\gamma = 0$ dB and $\lambda_a = 0.3$. Results are shown for different numbers of antennas.}
	\label{fig3}
\end{figure}

Figures~\ref{fig_cap} and \ref{fig3} plot the channel access probability (CAP) defined in Eqn.~(\ref{fdprocap}) and the successful transmission probability (STP) in Lemma~\ref{stplem}, respectively. Figure~\ref{fig_cap}(a) depicts the CAP versus node density. Increasing node density decreases CAP due to high contention among nodes in high density scenarios. As observed in Fig.~\ref{fig_cap}(b), with less sensitive PCS threshold $\Gamma = -30 $ dBm, more FD transmissions are likely to occupy the channel as opposed to a more conservative PCS threshold $\Gamma = -70$ dBm, which reduces the number of concurrent FD transmitters per time slot. A less sensitive PCS threshold value increases the number of concurrent transmissions and consequently, high interference is observed. This behavior of the channel access protocol necessitates the need for efficient PCS threshold selection. Figs.~\ref{fig3}(a) and (b) depict the STP versus SINR and node density, respectively, for different numbers of antennas. The STP is much lower at the high SINR of $20$ dB compared with the low SINR regime (e.g. $-20$ dB) due to high interference in large-scale networks. This is compensated for using multi-antenna transmissions. Similarly, at high node density $\lambda = 0.9$, high SINR regime is difficult to achieve due to increased numbers of concurrent transmissions generating high interference.

\begin{figure}[!h]
	\centering
	\includegraphics[width= 5in, height=3in]{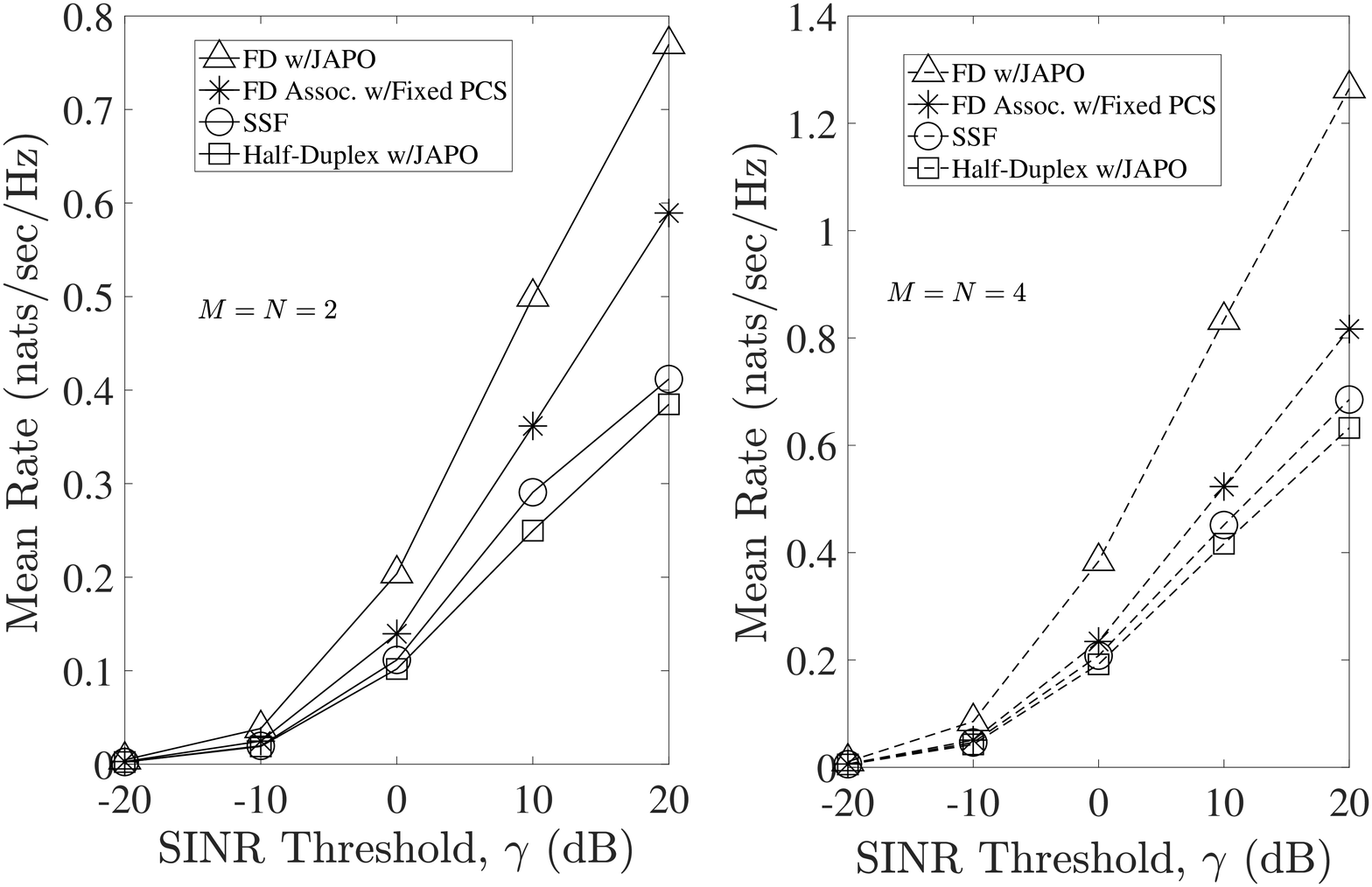}
	\caption{Mean rate versus SINR threshold $\gamma$ for node density $\lambda_s = 0.9$, $\lambda_a = 0.3$, $M = N = 2$ and $M = N = 2$.}
	\label{fig4}
\end{figure}

\begin{figure}[!h]
	\centering
	\includegraphics[width= 4in, height=3in]{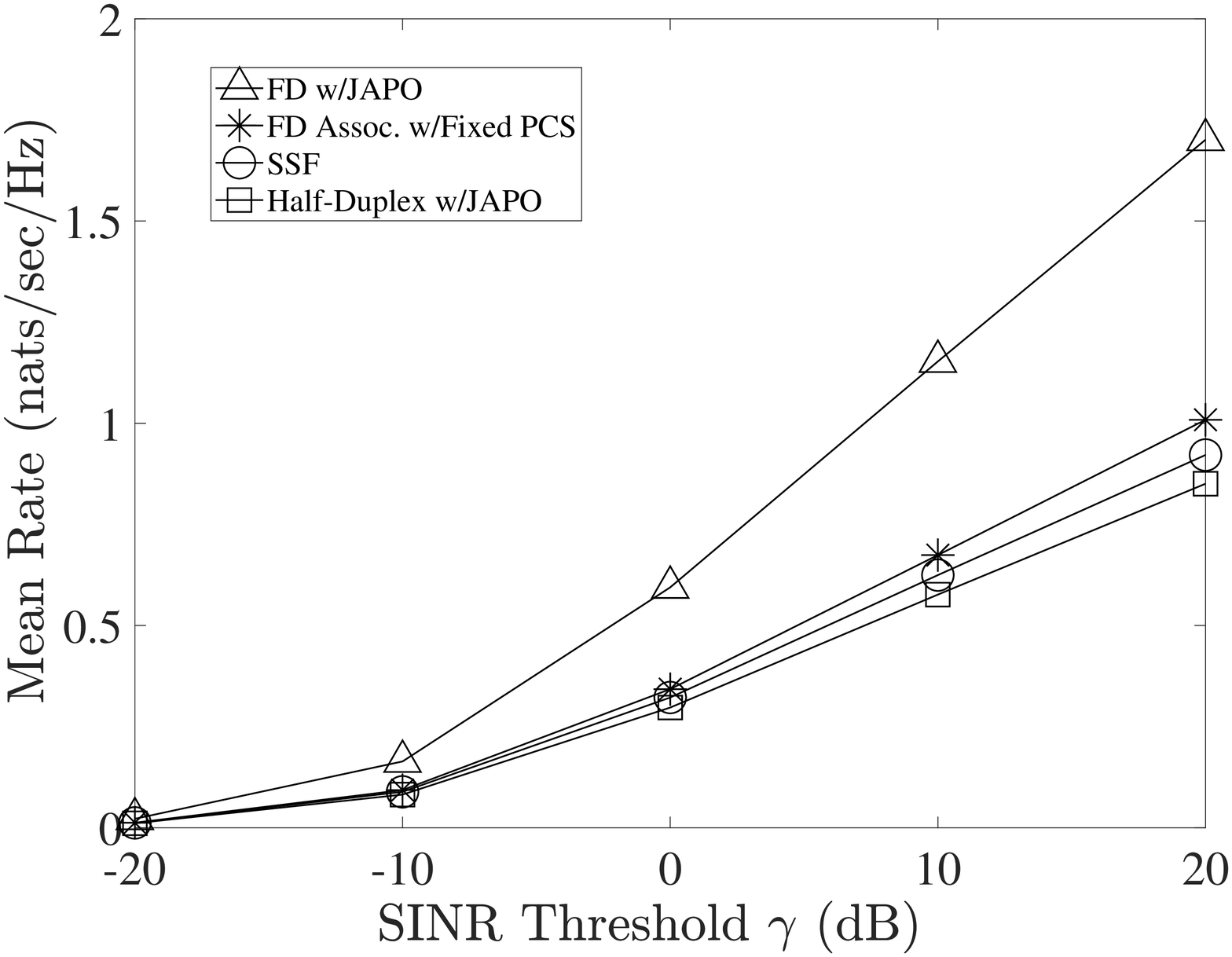}
	\caption{Mean rate versus SINR thresholds for $M = N = 8$, node density $\lambda_s = 0.9$ and $\lambda_a = 0.3$.}
	\label{fig5}
\end{figure}

In the presence high interference in large-scale networks, Figure~\ref{fig4} shows the performance gains at high STA density $\lambda_s = 0.9$ and fixed AP density $\lambda_a = 0.3$. Observing Fig.~\ref{fig4} at SINR $\gamma = 0$ dB and $M = N = 2$, the proposed algorithm JAPO doubles the mean rate ($0.2$ nats/sec/Hz) over the FD association without PCS threshold optimization. The AP association optimization with fixed PCS threshold offers performance gains over the existing SSF scheme, and in all cases, the mean rate is improved with multiple antennas. Fig.~\ref{fig5} shows the mean rate versus SINR for $M = N = 8$ and $\lambda_s = 0.9$. Under the FD Association with PCS threshold, the mean rate improves at high SINR and by jointly optimizing the AP association with PCS threshold, a further improvement is achievable. This additional gain is possible by optimizing the PCS threshold to guarantee that multiple concurrent transmissions are well separated in space to reduce the interference level in the network. 

\begin{figure}[!h]
	\centering
	\includegraphics[width=4in, height=3in]{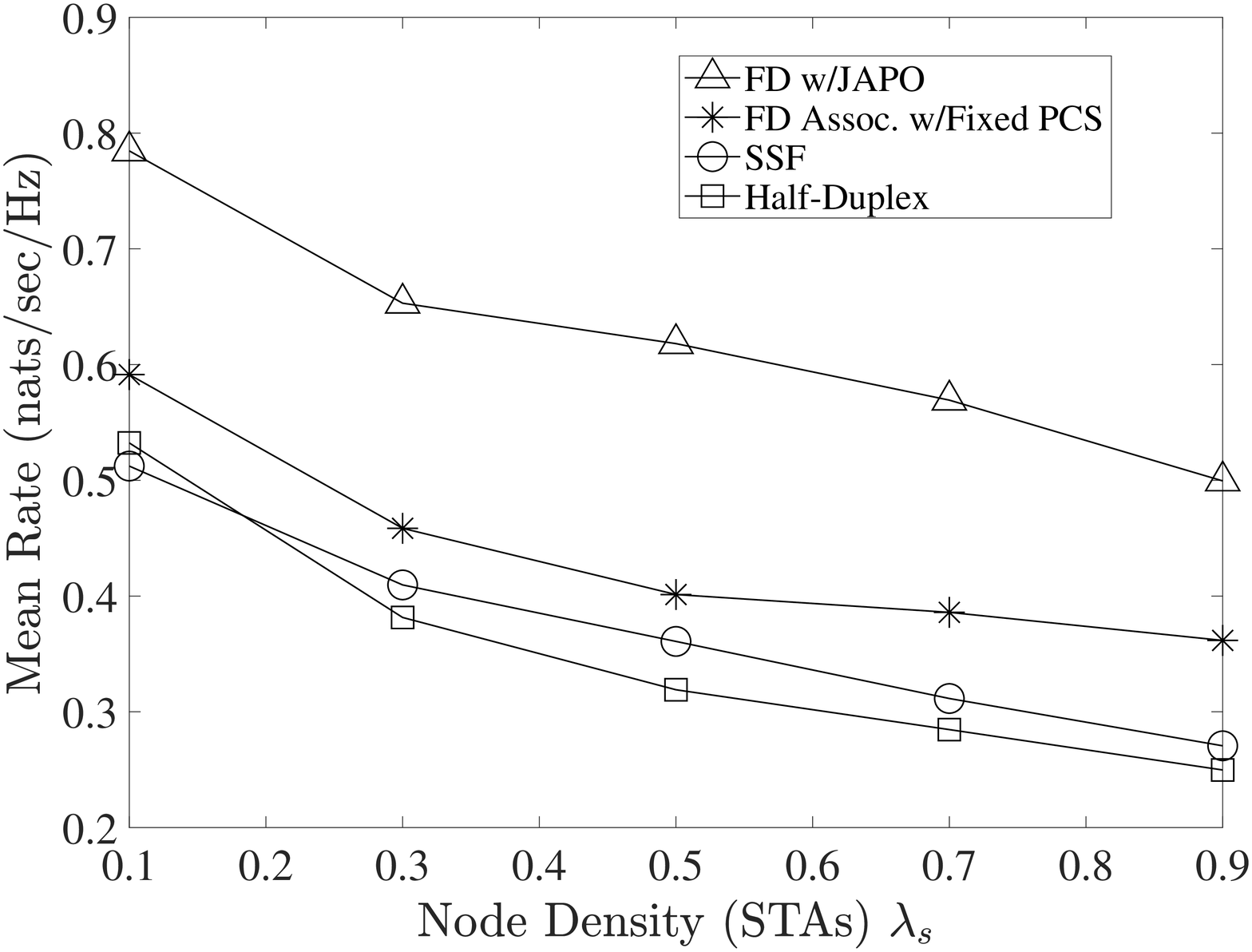}
	\caption{Mean rate versus node density $\lambda_s$ for user association with fixed PCS threshold and joint association and PCS threshold optimization given SINR threshold $\gamma = 10$ dB, $M = 2$ and $\lambda_a = 0.3$.}
	\label{fig6}
\end{figure}

\begin{figure}[!h]
	\centering
	\includegraphics[width=4in, height=3in]{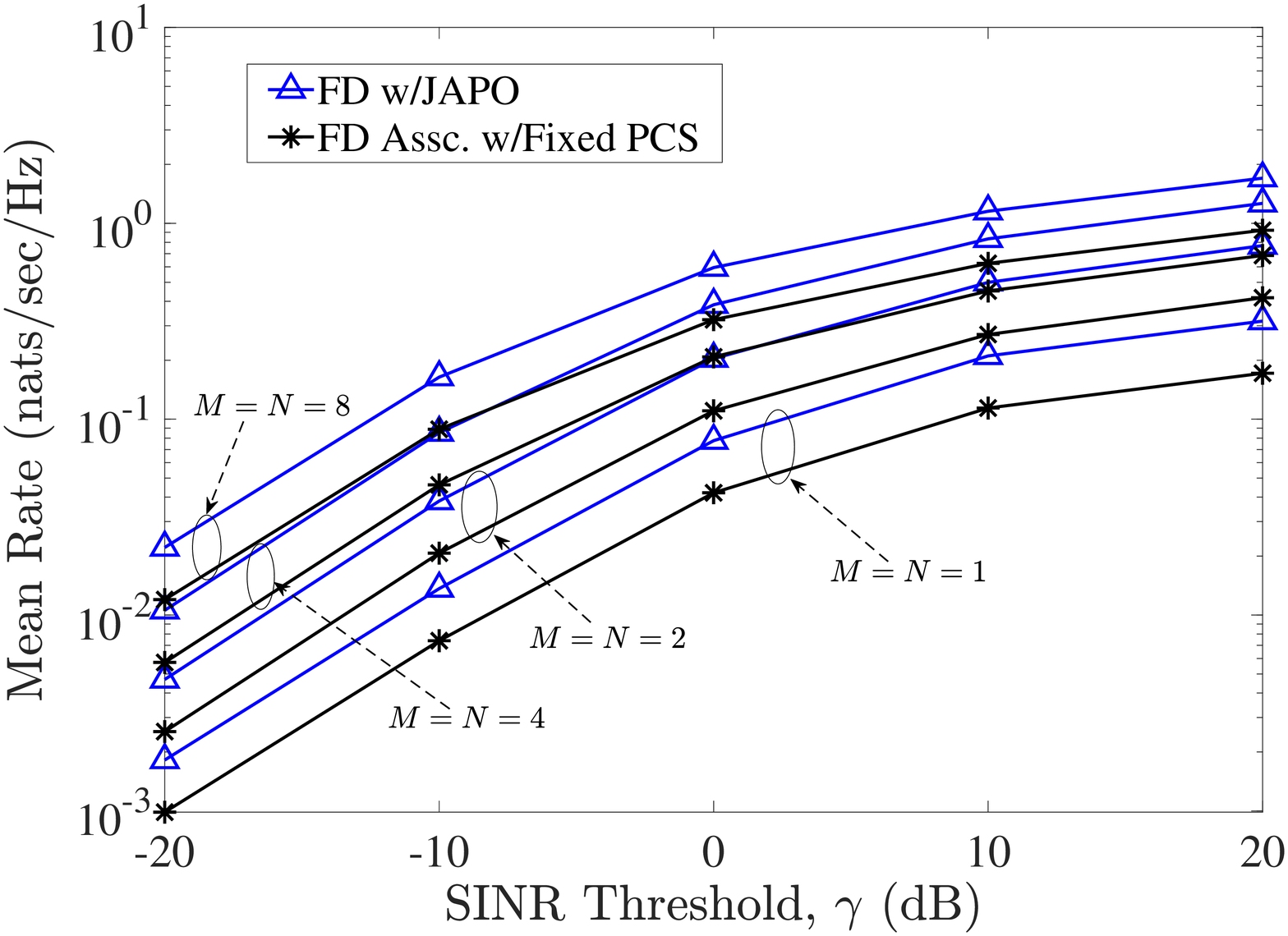}
	\caption{Mean rate versus SINR threshold for user association with fixed PCS threshold and joint association and PCS threshold optimization for various antenna sizes, node density $\lambda_s = 0.9$ and $\lambda_a = 0.3$.}
	\label{fig7}
\end{figure}

As shown in Fig.~\ref{fig6}, increasing to high node density $\lambda_s = 0.9$ is detrimental to the overall system performance because interference and contention tend to be more severe as node density increases. However, the proposed joint AP association and PCS threshold framework offers improvement in performance for mid to high node density over the case of optimizing only AP association and no AP association optimization. Taking a node density $\lambda_s = 0.9$ for example, an additional gain of 0.25 nats/sec/Hz is obtained over the AP association with fixed PCS threshold. Lastly, Fig.~\ref{fig7} compares the case of association optimization and joint association with PCS threshold optimization for various numbers of antennas. As shown, with increasing numbers of antennas from $M = N = 1$ to $M = N = 8$, the joint optimization framework further improves performance over AP association optimization with fixed PCS threshold $\Gamma = -70$ dBm, which might not efficiently control interference and contention.

\section{Conclusions}\label{conclusion}

Jointly optimizing the user-AP association and PCS threshold selection in high density wireless networks is proposed and its performance is assessed. Adding PCS threshold optimization to the AP association framework further improves performance significantly. The new proposed scheme jointly solves the user-AP association and PCS threshold selection problems assuming full duplex (FD) MIMO WLANs in the presence of out-of-cell interference and self-interference of FD transmissions. Though the schemes are suboptimal, performance evaluation reveals that spatial average throughput is improved by $24.4\%$ via AP association optimization alone. By combining AP association with PCS threshold optimization, a total throughput gain of $71.7\%$ is achieved for high node density. This additional $47.3\%$ gain is achievable by further optimizing the PCS threshold subsequent to AP association optimization. The key observation is that optimizing AP association yields performance gains for low to high node density in large-scale wireless networks. However, optimizing the PCS threshold jointly with AP association significantly further improves performance. In summary, the proposed joint AP association and PCS threshold selection framework is shown to be effective in achieving improved performance in high density networks where contention and interference are inevitable.




\end{document}